\documentclass[manuscript,screen,review=false]{acmart}

\AtBeginDocument{%
  \providecommand\BibTeX{{%
    \normalfont B\kern-0.5em{\scshape i\kern-0.25em b}\kern-0.8em\TeX}}}

\setcopyright{acmcopyright}
\copyrightyear{2025}
\acmYear{2025}
\acmDOI{XXXXXXX.XXXXXXX}

\usepackage{array,multirow}
\usepackage{cleveref}

\begin{document}

\title{Multiple Approaches for Teaching Responsible Computing}

\author{Stacy A. Doore}
\email{sadoore@colby.edu}
\affiliation{%
  \institution{Colby College}
  \city{Waterville}
  \state{ME}
  \country{USA}
  \postcode{04901}
}

\author{Michelle Trim}
\email{mtrim@cs.umass.edu}
\affiliation{%
  \institution{University of Massachusetts Amherst}
  \city{Amherst}
  \state{MA}
  \country{USA}
  \postcode{01002}
}

\author{Joycelyn Streator}
\email{joycelyn@mozillafoundation.org}
\affiliation{%
  \institution{Prairie View A\&M University and Mozilla Foundation}
  \city{Prairie View}
  \state{TX}
  \country{USA}
  \postcode{77446}
}

\author{Richard L. Blumenthal}
\email{rblument@regis.edu}
\affiliation{%
  \institution{Regis University}
  \city{Denver}
  \state{CO}
  \country{USA}
  \postcode{80221}
}

\author{Atri Rudra}
\email{atri@buffalo.edu}
\affiliation{%
  \institution{University at Buffalo}
  \city{Buffalo}
  \state{NY}
  \country{USA}
  \postcode{14260}
}

\author{Robert B. Schnabel}
\email{Robert.Schnabel@colorado.edu}
\affiliation{%
  \institution{University of Colorado Boulder}
  \city{Boulder}
  \state{CO}
  \country{USA}
  \postcode{80309}
}

\renewcommand{\shortauthors}{Doore, Trim et al.}

\begin{abstract}
Teaching applied ethics in computer science (and computing in general) has shifted from a perspective of teaching about professional codes of conduct and an emphasis on risk management towards a broader understanding of the impacts of computing on humanity and the environment and the principles and practices of responsible computing. This shift has produced a diversity of approaches for integrating responsible computing instruction into core computer science knowledge areas and for an expansion of dedicated courses focused on computing ethics. There is an increased recognition that students need intentional and consistent opportunities throughout their computer science education to develop the critical thinking, analytical reasoning, and cultural competency skills to understand their roles and professional duties in the responsible design, implementation, and management of complex socio-technological systems. Therefore, computing programs are re-evaluating the ways in which students learn to identify and assess the impact of computing on individuals, communities, and societies along with other critical professional skills such as effective communication, workplace conduct, and regulatory responsibilities.

 One of the primary shifts in the approach to teaching computing ethics comes from research in the social sciences and humanities. This position is grounded in the idea that all computing artifacts, projects, tools, and products are situated within a set of ideas, attitudes, goals, and cultural norms. This means that all computing endeavors have embedded within them a set of values. Through teaching students critical analysis methods, we can help them to identify potential biases, flaws, and unintentional harms in applications or systems if they can examine the underlying assumptions driving those designs and work with others to correct them. This kind of analysis makes space to bring real world technologies, stakeholders, and domain experts into the classroom for discussion, avoiding the pitfall of only engaging in toy problems. To teach responsible computing always requires us to first recognize that computing happens in a context that is shaped by cultural values, including our own professional culture and values.   

The purpose of this paper is to highlight current scholarship, principles, and practices in the teaching of responsible computing in undergraduate computer science settings. The paper is organized around four primary sections: 1) a high-level rationale for the adoption of different pedagogical approaches based on program context and course learning goals, 2) a brief survey of responsible computing pedagogical approaches; 3) illustrative examples of how topics within the CS 2023 Social, Ethical, and Professional (SEP) knowledge area can be implemented and assessed across the broad spectrum of undergraduate computing courses; and 4) links to examples of current best practices, tools, and resources for faculty to build responsible computing teaching into their specific instructional settings and CS2023 knowledge areas. 

\end{abstract}



\received{XXXX}
\received[revised]{XXXX}
\received[accepted]{XXXXX}

\maketitle

\section{Introduction}
This paper presents readers with guidance and methods of teaching computing ethics from multiple epistemological (ways of knowing what we know or believe to be true)  perspectives.  This text enacts an awareness of the roles that power and identity play in research, education, and workplace situations in more depth than may be customary in more traditional approaches to ethics education. Our hope is that this resource will provide guidance and suggestions that readers will find adaptable for many courses, contexts, and lessons.  In creating this resource, we want to contribute different ways of understanding what computing ethics is, what sorts of questions it should include, and what aspects of human activity we believe it must consider. One goal of this opening section is to make explicit some of the assumptions and epistemologies driving our work and to invite  practitioners to take the ideas presented here further as they develop their teaching methods, modules, activities, and classes.

\textit{Why teach ethics in computing?} In exposing our assumptions, we begin with this question in order to clarify arguments and motivations for integrating ethics into computing education. Whenever practices, such as medical care or education, or technologies, such as planes, bridges, and water wells, become part of the fabric of public life, there exists the potential for injury, lasting harm and even death to the humans using those services and tools. Computing has become a practice, a way of acting on society, and it has become a collection of tools with varying degrees of potential for causing harm to human beings. As computing expands into more domains and into more areas of the public sphere, that potential for harm exponentially increases. While there is great potential for good to come from computing as it infiltrates more areas of society, computer science as a relatively new discipline has grown faster than it has had time to reflect on how computing impacts the world. Development cycles are increasingly shorter, and inadequately tested systems and tools are reaching users and impacting communities with significant ethical debt. Educators have the responsibility to teach ethics so that computing students bring ethical reasoning, values, reflection, and responsible practices to their design and development activities. This integration is particularly important in the sciences because the scientific method which assumes repeatability and objectivity does not on its own convey a system for assessing how cultural forces and systems of oppression can intersect with particular computing practices to amplify injustice and inequity. In addition to the moral argument for taking societal considerations into account while designing computing systems and hence the need to teach it, we think this pursuit is intellectually rigorous in its own right. Indeed, constraints introduced by societal considerations can suggest new and interesting problems in the traditional computing space  (though we hasten to add that in some cases the pursuit of a pure technological solution can be very misguided).

\subsection{Teaching through a Multiple Perspectives Lens}
The myth of objectivity, otherwise referred to as objectivism or objective theory, can get in the way of making ethics instruction relevant for contemporary computer science. Because the world of computing design, development, and implementation is incredibly diverse, there must be acceptance for more than one way of knowing and understanding the world.  When approaching a situation where a decision-maker seeks to implement a computing solution, there are always multiple perspectives identifying which problem needs to be addressed and the extent to which a computing solution should be applied.  Disciplinary training and cultural background shape our method for sense-making. As we learn where to focus our attention in approaching a problem, we also acquire blind spots. A seemingly mundane example of this sort of blindspot is the discovery that light reflective activating hand dryers, faucets, and soap dispensers used in many public restrooms are unable to ‘see’ darker skin tones \cite{benjamin2023race}. In this case, for those with lighter skin, that lack of melanin is a blind spot. In other words, light skin is a feature  assumed to be universal, making whiteness or light-skinned-ness, the default standard without designers consciously thinking about it. As frustrating as not being able to wash one’s hands can be, the inability for many digital cameras to accurately recognize darker skin toned faces extends to inaccuracy in facial recognition algorithms and thus to all of the civic processes built on that technology. \cite{buolamwini2018gender, najibi2020racial}. Even oximeters, those finger devices in hospitals that measure oxygen saturation, can inaccurately report values for dark skinned patients, possibly placing Black patients at increased risk for hypoxemia \cite{sjoding2020racial}. These blind spots, or \textit{unknown unknowns}, are why issues of identity, representation, power, and access must be considered in exploring or examining the ethical issues of a particular computing action or design decision.  Lawmakers and the public are expecting algorithmic decision systems to be fair, resulting in companies making the difficult decision to shelve projects that don’t work as intended \cite{rivero2020}. All computing artifacts, projects, tools, and products are situated within a set of ideas, attitudes, goals, and cultural norms. This means that all computing endeavors have embedded within them a set of values. Through critical analysis, we can reveal potential blind spots in application or tool or system design if we work to identify the underlying assumptions driving those designs.  This kind of analysis makes space to bring real world technologies into the classroom for discussion, avoiding the pitfall of only engaging in toy problems. To discuss ethics in computing is always to first recognize that computing happens in a context that is shaped by cultural values.   

\paragraph{Overview of the paper.} In \Cref{sec:terms} we define concepts and terms related to responsible computing that we use throughout the paper. \Cref{sec:approaches} considers various approaches to teaching ethics and responsible computing. We present our recommendations in \Cref{sec:recommend} and some example responsible computing interventions in \Cref{sec:examples}. Finally, we conclude in \Cref{sec:concl}.

\section{Computing Ethics and Responsible Computing: Concepts and Terms}
\label{sec:terms}

\textit{What do we mean when we say ethics in computer science?} Many of the instructors teaching in computer science or engineering departments in this first quarter of the 21st century might have encountered ethics at some point in their education.  This exposure may have manifested in a number of likely ways: via lessons on professionalism, case studies centered on catastrophes or debates on the morality of specific, hypothetical situations. In the first method for teaching ethics, emphasis might be placed on honesty, intellectual property rules, or in avoiding conflicts of interest. In the second,  instructors might share with students a case: a detailed, chronologically narrativized description of an unfortunate event, such as a crash, breakdown, or critical systems failure. Then, that case will be analyzed using a post-mortem style investigation, usually with the goal of leading students to discover and explain what went wrong, i.e., which pivotal action or set of actions led to the catastrophe. In hypothetical debates about morality, students are often presented with a conundrum with no obvious good outcome to raise the stakes for exploring the morality of the invented issue at hand. The trolley problem is a classic example of such a conundrum. In this example, a trolley is speeding toward a switch or a fork in the track. Directly ahead is an elderly person or a child and on the other track is a group of people. The morality debate comes from advocating for which person or persons should the driver choose to run over (see for example the Moral Machine \cite{moralmachine}).  While all three of these traditional approaches have their merits, they may not be sufficient for equipping today’s students with the knowledge and analytical practice needed to negotiate the ethical challenges presented by ubiquitous computing in a heterogeneous global context.

A shift from ‘ethics’ toward ‘responsible computing’ within the larger computer science community seems to signal a recognition that traditional approaches to what have historically been called ‘ethics’ education may need an update.  Recognizing some of the limitations of traditional case study approaches, initiatives in some institutions have sought inclusions from classical philosophy. With its common focus on logic and rationality, and as the disciplinary home of the study of Ethics, the inclusion of philosophical approaches to ethics within computer science education might be sensible. Regardless of the intellectual tradition instructors draw from, it is important to recognize that there are multiple intellectual traditions, and a one-size-fits-all approach that prioritizes one epistemology may not equip students well for the diverse problems facing our world today.  While trolley problems can seem fun to debate, they typically do not consider dynamics of power and identity that can shape the range of any human’s perceived options for available actions in a given situation. Case studies can be useful tools for expanding students’ sense of what’s possible. It is not clear to what extent students learn how to recognize when they are on the cusp of an action or decision that has ethical implications based on lessons relying on case studies alone. Students experience a more robust ethics education when they combine discussion, analysis, application, and reflection.  This multi-faceted pedagogical approach resists the pitfalls that can occur when a lesson or case assumes a clear and distinct and fully knowable set of circumstances from one perspective and with predictable outcomes.  In other words, an instructor can do students a disservice by not addressing the possibility that an unexamined assumption can introduce harmful bias into a design process or that developers can follow the ‘rules’ and still contribute to negative impacts. 

Taking a cue from decolonial design practices \cite{smith2021decolonizing}, an instructor can allow for other ways of knowing and understanding the world in their class, such as prioritizing human dignity or user privacy over efficiency. Exposing students to multiple ethical frameworks and theories invites them to locate their own lived experiences and personal ethical codes within their computer science problem solving and research activities.  This diversity of perspective and integration of reflective practices provides opportunities for students to examine and become cognizant of their own implicit biases and assumptions. In approaching teaching ethics in a computing classroom, instructors might consider how they can help students to first identify what assumptions are operating. For example:
\begin{itemize}
    \item What is the goal of a given project or analysis? 
    \item Should a design be accessible and inclusive?
    \item Should it be sustainable? 
    \item Should it be cost effective and efficient? 
    \item Should it adhere to standards for data privacy and data accuracy?
    \item Should it facilitate vulnerable populations’ safety? 
    \item What are the ethical values operating in a given context? 
    \item Does the problem really need a computing/technology solution?    
\end{itemize}
Making those values part of projects, assignments, and examples is one way to bring into the room what ethical design should look like. In addition, having conversations and inviting reflections from students about how those ethical principles manifest in a design brings home this connection between values and decisions.

\section{Approaches to Teaching Computing Ethics and Responsible Computing}
\label{sec:approaches}

This section provides a brief review of the literature around two distinct approaches to teaching computing ethics at the undergraduate level: 1) a dedicated, intermediate level Society, Ethics and Professionalism (SEP) focused course, and 2) an SEP curriculum embedded into CS core courses at different granularity levels (e.g.,  a single lesson/assignment, self-contained module, an integrated module, and an SEP theme woven throughout the course). We will briefly review each of these approaches, their advantages and challenges, and the ways in which they offer students different perspectives on responsible computing principles and practices.

\subsection{Dedicated SEP Course Approach}
The traditional approach to teaching a society, ethics and professional responsibility and conduct focused course has been to offer the content as a stand-alone, intermediate level course fulfilling the required number of hours and topics to meet ABET accreditation and ACM/IEEE curricular standards \cite{raj2022interpreting,sahami2014computer}. Recent studies have focused on how ethical and social responsibility in computing are understood and used in instruction by computer science faculty \cite{fiesler2020we, smith2023incorporating, stavrakakis2021teaching, garrett2020more}. A review of syllabi from dedicated SEP courses documented trends in instructor discipline, content/topics, instructional approaches, and learning objectives \cite{fiesler2020we}. Overall, the study found that most stand-alone computing courses are taught within computer and information science departments and instructors of these courses are just as likely to have a degree in philosophy, science and technology studies, sociology, cultural studies, or information science as computer science. The review found the most frequent categories for course topics included law, policy and regulation, privacy, philosophical frameworks, social justice and inequality, algorithmic impact on society and the environment. Most course learning objectives were related to identifying ethical issues in real world scenarios or case studies, critically evaluating the identified issues from multiple perspectives, communicating effective arguments in writing, and applying rules such as policies, codes of ethics, and philosophical frameworks to develop potential solutions to the identified issues. \cite{fiesler2020we} 

A survey of computer science departments across the European Union (EU) found that a third of the departments were not teaching computing ethics in their programs and those that did devoted a small percentage of time (on average >10 hours per program) and most often as a stand-alone class \cite {stavrakakis2021teaching, blumenthal2021walking} conducted a similar study through a document review of graduation requirements specified in the Academic Catalog of 500 Bachelor of Science in Computer Science (BSCS) programs in the United States. This survey found that 45 percent of these BSCS programs required students to take a single required SEP focused course that was identified as residing in the computer science department. In unreported data from the same review survey, an additional 7 percent of programs required computer science students to take an SEP technology focused course (as opposed to a general ethics course) identified as residing in the Philosophy department. As a result, half (52\%) of all BSCS programs in this survey required students to take at least one technology-focused SEP course as part of their computer science curricular requirements.

\subsubsection{Barriers to Expanding Computing Ethics Instruction}
Although there is currently much interest in materials and approaches to improving the quality of computing ethics instruction in both dedicated and embedded courses, there are persistent barriers that prevent many CS faculty from implementing new practices and techniques. A recent study \cite{greer2020overcoming} conducted qualitative interviews with CS department chairs and faculty to better understand barriers to moving from a dedicated course approach to an embedded ethics across the CS curriculum approach. Their findings suggest that traditional computing course faculty often omit discussions of computing ethics because of a lack of expertise, perceived time constraints, or lack of appropriate teaching materials. There was also a considerable deference to authority (i.e., ABET accreditation, NAE, NAS) given as the reason for not devoting more department time and resources to helping faculty acquire the expertise needed to embed responsible computing instruction throughout the technical curriculum.

\subsubsection{Dedicated Computing Ethics Approach}
There is a history of high quality, dedicated computing ethics courses that provide an opportunity for students to engage deeply in topics such as technology creation issues such as privacy security, equity, and access through creative and innovative assignments and materials \cite{johnson1994computer, sanders2005discussion, ferreira2020computer, reich2020teaching}. More recently interdisciplinary initiatives are using narrative forms of science fiction, film, and theater to engage computer science students to reason about potential impacts of past, present or speculative future technologies on humans and on society \cite{skirpan2018quantified, burton2018teach, burton2023computing, doore2022release, fiesler2021integrating, shapiro2021using}. 

The advantages of an intermediate level dedicated course come with the time and flexibility of having a whole semester to help students explore issues in an in-depth manner. Rather than have the course be a student’s only exposure to these topics, ideally, the course should be a part of a holistic approach to developing students’ knowledge and skills in responsible computing principles and practices over the entire course of their computer science program. For example, a first year student might participate in a short embedded lesson on the importance of verifying data set provenance and validation of data accuracy in a traditional introduction to computing course while learning how to clean .csv and import demographic data before conducting a data analysis lesson in Python. When the student has advanced to an intermediate level, they may enroll in a dedicated course where they spend a few weeks of class time looking at individual and societal level impact of dataset preparation such as data collection, datasheet construction, pilot model building, testing and validation, and monitoring/auditing responsibilities as well as talking about the assumptions that are inherent in the data preparation pipeline. While focusing on the skills of dataset preparation, a dedicated course can examine questions of power, inequality, and critical ethical issues with real world data collection practices (e.g., web scraping, involuntary surveillance, informed consent, compensation, politics of classification, etc.) and how this impacts the conclusions and consequences that result from dataset preparation processes. It can provide extended opportunities for students to do research and report on current best practices in dataset preparation with identified stakeholder interests, questions and checkpoints embedded into the dataset lifecycle, and the ethical and professional responsibilities of a technologist before a dataset is ever used for a ML application or AI system. Finally, these same students will have the training and the tools to apply their in-depth knowledge of responsible computing principles and practices to capstone projects, independent studies, and honors theses, which would require these components in final presentations and research products. A dedicated course provides time to invite faculty with disciplinary expertise into computing courses to provide historical and political context for inequalities and power differentials in society and how technologies have supported existing structures.

Despite the many advantages that come with a dedicated course structure, there are known issues with the ‘stand alone’ model of teaching computing ethics through offering only a single course targeting 3rd or 4th year students. There is a concern about the quality of these single courses taught by computing faculty without a humanities or social science background and whether it is sufficient to provide only a brief overview of applied ethical theory and its relation to the impact of technology \cite{brey2000disclosive, brey2000method, goetz2023}. While the dedicated single course approach offers a simple model for rotating CS faculty to teach, the result is often a lack of depth in the complex and fluid relationships between innovations in computing and the societal context and power structures such as class, race, gender, and ability \cite{mehdiabadi2019facing, smith2023incorporating}. Another identified potential weakness with the single stand-alone course approach is that it can often leave students thinking that the content is not central to their computer science degree because of the disconnect from the core technical curriculum topics. Studies on student attitudes towards computing ethics suggest students fail to see the relevance of dedicated courses when that is the only instruction provided in their program and the only time they see their instructors raising these issues \cite{califf2005effective, spradling2008ethics, davis2011incorporating, cech2014culture, goldweber2022hands}

\subsection{Embedded Computing Ethics}
Within the past ten years, there has been a growing recognition that many students do not have sufficient opportunities to connect the training in traditional computer science concepts and skills with the impact of rapid and continuous computing advances on individuals, communities, society, or the environment. Despite a series of accreditation standards that require programs to provide a minimum number of instructional hours devoted to professional conduct and ethics \cite{raj2022interpreting,abet-cac}, there is enough evidence in academia and industry settings to suggest the single dedicated course approach does not provide enough training for computing professionals at all career stages to critically examine their responsibilities as creators of technologies nor the consequences of their designs, actions, and impact on anyone else other than themselves and their future career paths.

In response, there is a growing movement of computer science educators and industry partners who have been developing instructional materials on responsible computing principles and practices that are intentionally interdisciplinary and embedded within the core computing curriculum areas \cite{grosz2019embedded}. These materials come in many different formats but span the range of instructional levels from introduction to programming (CS00/01) to advanced computing concepts (CS03/04). There are now a variety of resources available to help faculty get started in integrating responsible computing into their courses and will answer questions related to administrative and colleague support through to the effective assessment of the lessons. Many of these resources have been highlighted in the National Academies of Science (NAS) Fostering Responsible Computing Research report \cite{national2022fostering} but there are many more resources being released every month. These materials come in many different formats but span the range of instructional levels \cite{cohen2021new, fisler2022approaches, mcdonald2022responsible, zuber2022empowered} and open educational repositories of these materials are widely available \cite{embedethicsharvardrepo, computingnarrativesrepo, engagecseduethicsrepo}. There are also a variety of resources available to help faculty get started in integrating responsible computing into their courses and will answer questions related to administrative and colleague support \cite{rcscplaybookoverview2020} through to the effective assessment of the lessons \cite{horton2022embedding, horton2023more,bemidjirepo, ethics4eu, peckethics}. In the next section, we provide a description of different categories of embedded lesson formats with examples of each type, along with an example of a lesson that might be found in a dedicated semester-long course.

\section{Recommended Best Practices}
\label{sec:recommend}

We present some best practices in incorporating ethics (and responsible computing more generally) into computing as follows. In \Cref{sec:intervention}, we first revisit the taxonomy of responsible computing interventions from the previous section and present some recommendations of which interventions an instructor should consider in their course based on how much preparation time the instructor has as well as their comfort/experience with responsible computing interventions. Once the intervention type and course is picked, we present some best practices for what to do (i) before a course has started (in \Cref{sec:before}), (ii) during the course (in \Cref{sec:during}) and (iii) after the course is completed (in \Cref{sec:after}). For additional issues and recommendations, please see the Mozilla Teaching Responsible Computing playbook~\cite{rcscplaybookoverview2020}.

\subsection{Choosing an Intervention Approach}
\label{sec:intervention}
In the following subsections, we present five different interventions for responsible computing material along with some preliminary recommendations. We have roughly ordered the interventions in order of time commitment needed to create (or incorporate from existing resources) them into a course. The first four fall in the category of `embedded ethics' where the idea is to incorporate responsible computing thinking into traditional computing courses (e.g. CS I, II, III, Algorithms etc.) while the last one is having a dedicated course on responsible computing.

\subsubsection{Individual Assignments or Problems}
\label{sec:indiv-assignment} The idea for such interventions is to use responsible computing to provide the motivation or a `story' for an assignment or problem in a traditional computing course. The actual solution of the problem would essentially be as in an earlier incarnation of the course (e.g. students still build a specific data structure) so the assignment/problem will still test whether the student is able to a grasp a traditional computing concept but now applied in a different context from what they have seen in class. (See \Cref{sec:assess} for more on this.) See \Cref{tab:indv-prob-ex} in the appendix for illustrative examples.

Among all the interventions included in this review of approaches, this one most likely will need the least amount of preparation. For instructors who have not incorporated responsible computing in their classes before or who are short on preparation time, this might be the easiest intervention to start off with and we recommend this intervention over the others.

\subsubsection{Self-contained Lesson/Module}
\label{sec:self-contained-module}
The idea in this intervention is to create a self-contained lesson plan/module that talks about responsible computing in a traditional computing course but the module is somewhat disconnected from the material covered in the rest of the course (though the module fits into the overall course topic). See \Cref{tab:self-cont-module-ex} in the appendix for illustrative examples.

If possible, we recommend creating an integrated lesson/module. However, a self-contained module might be easier to develop if the module is led by and/or created by someone other than the instructor. The idea here is to bring an outside expert (e.g.,  philosopher, social scientist, or critical studies researcher) to come to class and lead the discussion on responsible computing in a scenario that is relevant to the course but from a social impact perspective.

\subsubsection{Integrated Lesson/Module}
\label{sec:integrated-module}
The idea in this intervention is similar to the earlier section but the lesson plan/module covers a topic that had already been covered in the traditional computing class prior to the current intervention. See \Cref{tab:int-module-ex} in the appendix for illustrative examples.

We recommend this intervention over the self-contained module because having a module that is tightly integrated into the course material as there is growing evidence this may be beneficial \cite{horton2022embedding} for the students over both the short and longer term. The instructor of the course must develop (or co-develop) this module to ensure that the integration with the rest of the course material is seamless.

\subsubsection{Responsible Computing Theme }
\label{sec:RC-theme}
This intervention is the most extensive among the ones we present for a traditional CS course. The idea here is to pick a theme (e.g. access to high speed Internet) and use that theme across multiple assignments/lectures/projects in the course.  See \Cref{tab:rc-theme-ex} in the appendix for illustrative examples.

\subsubsection{Dedicated Course}
\label{sec:dedicated}
This intervention is an entire course that is dedicated to a series of subtopics in responsible computing. One of the goals of such a dedicated course is to allow space for longer and more nuanced discussions on issues at the intersection of society in computing, which due to time constraints might not be possible in a traditional course. See \Cref{tab:dedicated-ex} in the appendix for illustrative examples.

We do not recommend the dedicated course intervention over the other four interventions in a traditional computing course \textit{or} vice-versa. We recommend that a department implements both a dedicated course as well as interventions in traditional computing courses.

\subsubsection{Combined Intervention Types}
\label{sec:combined-types}

In many ways, this would a sixth type of intervention - do them all - but most programs might need substantial time and curriculum development to reach this stage so this might be considered a long-term program goal rather than a place to start. If this is a program goal, it would be best to look for programs that have been doing this work for some time to see how they built new courses into their schedule and revised previous courses to integrate responsible computing into their core curriculum (see \Cref{sec:resources} for some examples).
\newcommand{\DC}{DC}
\newcommand{\PC}{PC}
\newcommand{\NNC}{NNC}

\newcolumntype{P}[1]{>{\centering\arraybackslash}p{#1\textwidth}}

\begin{table}[]

    \centering
    {\renewcommand{\arraystretch}{1.2}%
    \begin{tabular}{|P{0.13}|P{0.124}|P{0.124}|P{0.124}|P{0.124}|P{0.124}|P{0.124}|P{0.124}|}
    \hline
         &  \multicolumn{6}{c|}{\textbf{Curriculum Development Stage}}\\ \cline{2-7}
         \textbf{ Curriculum Dimensions}& Individual Assignments & Self Contained Module  &Integrated Module  & Responsible Computing Theme  & Dedicated Course  & Combined Intervention Types \\ 
         & (Sec.   \ref{sec:indiv-assignment}) &  (Sec. \ref{sec:self-contained-module}) &(Sec. \ref{sec:integrated-module}) & (Sec. \ref{sec:RC-theme}) & (Sec. \ref{sec:dedicated}) & (Sec. \ref{sec:combined-types})\\         
         \hline
    Using Existing Resources (Sec. \ref{sec:existing-rsources}) & \DC & \DC &\NNC &\NNC &\NNC&\NNC\\
    \hline
    Student Knowledge/Experience (Sec. \ref{sec:student-prior}) & \DC & \DC &\DC &\DC &\DC&\DC\\
    \hline
    Responsible Computing Community (Sec. \ref{sec:rcc}) & \PC & \DC &\DC &\DC &\DC&\DC\\
    \hline
    Social Justice (Sec. \ref{sec:sj}) & \PC & \DC &\DC &\DC &\DC&\DC\\
    \hline
    Learning Objectives (Sec. \ref{sec:LO}, \ref{sec:assess})
 & \PC & \DC &\DC &\DC &\DC&\DC\\
    \hline
    \end{tabular}
    }
    \caption{Responsible Computing Curriculum Planning Considerations 
                 (\DC= Definitely Consider, \PC= Perhaps Consider, \NNC= Need Not Consider)}
    \label{tab:before-class}
\end{table}

\begin{table}[]
    \centering
    {\renewcommand{\arraystretch}{1.2}%
    \begin{tabular}{|P{0.13}|P{0.124}|P{0.124}|P{0.124}|P{0.124}|P{0.124}|P{0.124}|P{0.124}|}
    \hline
         &  \multicolumn{6}{c|}{\textbf{Curriculum Development Stage}}\\ \cline{2-7}
         \textbf{ Curriculum Dimensions}& Individual Assignments & Self Contained Module  &Integrated Module  & Responsible Computing Theme  & Dedicated Course  & Combined Intervention Types \\ 
         & (Sec.   \ref{sec:indiv-assignment}) &  (Sec. \ref{sec:self-contained-module}) &(Sec. \ref{sec:integrated-module}) & (Sec. \ref{sec:RC-theme}) & (Sec. \ref{sec:dedicated}) & (Sec. \ref{sec:combined-types})\\         
         \hline
    Difficult Conversations Section 
 (Sec. \ref{sec:difficult}) & \NNC & \NNC &\PC &\PC &\DC&\DC\\
    \hline
    Student Motivation Section (Sec. \ref{sec:student-motivate}) & \DC & \DC &\DC &\DC &\DC&\DC\\
    \hline
    Assessing Learning (Sec. \ref{sec:assess-learning}) & \PC & \DC &\DC &\DC &\DC&\DC\\
    \hline
    \end{tabular}
    }
    \caption{Responsible Computing Curriculum Implementation Considerations 
                 (\DC= Definitely Consider, \PC= Perhaps Consider, \NNC= Need Not Consider)}
    \label{tab:during-class}
\end{table}
\subsection{What to do before the lesson/course?}
\label{sec:before}

In this section, we present some questions  and issues that the instructor should consider while the intervention or course is being designed. Most of these issues should be considered in parallel instead of in sequence, hence we recommend ignoring the ordering here. \Cref{tab:before-class} provides our recommendations of which questions to consider when planning for specific interventions.

\subsubsection{Using Existing Resources vs. Creating Own Resources}
\label{sec:existing-rsources}
The answer depends on the time constraints as well as the comfort of the instructor. For interventions that are individual assignments/problems or self-contained modules, using an existing resource might be the easiest way to start. The somewhat isolated nature of these interventions would make porting existing resources easier than other interventions. The other obvious advantage is that using existing resources needs much less time commitment from instructors, which is a big advantage for encouraging adoption and adaptation \cite{smith2023incorporating}. (See \Cref{sec:resources} for links to Responsible Computing Repositories and Resources websites.)

The other three interventions are much more closely intertwined with the existing course structure and might be harder to port over. Having said that, if the traditional course is one of the standard lower level courses (e.g. CS-01/02/03), then porting over an integrated module or a theme might also be possible. Using existing material might be harder for elective courses where the topic coverage tends to vary a lot, akin to delivering a new lecture from someone else’s slides.

A dedicated course probably is best suited to be created by the instructor or co-instructors (or at least be modified from an existing course resource ). As we will see next, using examples and case studies (or even the overall course topic) that would resonate most with the intended student population means porting courses that are `fine-tuned' to one school might not work at a school with a substantially different student profile, course enrollment size, or instructional context.

\subsubsection{Student Prior Knowledge and Experience}
\label{sec:student-prior}
Since a lot of responsible computing intervention involves discussion about questions that do not have a definitive answer (unlike many traditional computing courses), it is important that students be engaged in these discussions. This means care must be taken that the topics picked for discussion will resonate with students: or at the very least the `level' of discussion that one can expect depends on where the students are in their life journey. If the instructor is using an existing resource then they should pay close attention to whether the resource is well suited for their students (or if the existing resource needs to be altered/adapted)-- again these might be less of an issue for interventions that are individual assignments/problems or self-contained modules.

\paragraph{Student Lived Experience}
The lived experience of a student and the cultural context of the institution will have a key role in how they will engage in the discussions. For example, a discussion about structural racism may be approached in a different way in an HBCU context than from a school where white students form the majority. Or a class with a sizable international student population will react to a discussion on racism in the US differently than a class with mostly domestic US students. Different regional cultures have different levels of cultural competency training integrated into elementary and secondary schools, and careful scaffolding may be required to foster a safe space in the classroom for all students. Needless to say, discussions reflecting the different makeup of students’ lived experiences can be rich learning opportunities but the instructor should be aware of how their students’ identities may impact the discussions so they can structure the lessons in a way that is most fruitful for all of their students.

\paragraph{Humanities and Liberal Arts Experience}
Many of the nuanced discussions mentioned earlier need the students to understand how the society functions and various existing structures and their existing comfort with such material. A class where all students have been exposed to Black Feminist theory will have a very different discussion than a class where students have not taken a single class on gender or race. In most cases, this depends on the culture of the institution and can vary greatly.

The reason why the above is important is that it determines how much scaffolding the instructor has to put in to make sure that the students participating in a discussion have a minimum amount of exposure to relevant background material. For example, if most of the students have not thought much about systematic racism in the US (maybe because of their lived experience e.g. international students or their (lack of) background in liberal arts education), then the instructor should spend some time going over some basics of racism (especially as it pertains to the US). These need not always be academic textbooks but could be other narratives (e.g., videos, podcasts, short literature, etc.). These lessons could be facilitated by guest lectures from a disciplinary expert in the humanities.

More generally, a lack of student training in the humanities means that (unlike programming where students have a common language) students will not have a common language to express their opinion. Thus, the instructor should think about structured ways to talk about social issues so that students have the conceptual tools and the appropriate vocabulary to articulate their thoughts in the discussions in ways that maintain a positive learning environment for all.

\paragraph{Student Apathy or Resistance}
A common issue, especially with students that do not have the lived experience or the proper humanities background is that while students would acknowledge for example the reality of systematic racism, they would be apathetic to discussion on race in a computing class. Sometimes this apathy is because the students think “this is not computing’s problem” or sometimes the apathy is a result of feeling helpless when confronted with such huge societal problems that are impossible to fix individually. They may also be feeling disempowered generally as sweeping changes in computing professions happen rapidly, often led by global companies with such a pervasive influence that students see no way they could possibly impact a given trajectory.  Irrespective of the reason behind the apathy, the existence of apathy would result in students `checking out' of discussions (which they could have really gained from if they participated more openly). On the other hand, providing opportunities for reflection enables those quiet students who are engaged to signal their engagement, as there will always be students who don’t tend to process their learning through talking in class. This leads to the next question.

\paragraph{Opening with the Responsible Computing `Hook'}
The instructor needs to decide on the `hook' that will be of interest to even students who might struggle to see how a particular issue impacts them. For example, instead of talking about race directly at the start of a class, perhaps discuss a topic like access to technology (as an even more specific example access to high speed Internet) that all students can relate to \cite{rcsplaybookaccess}. Instructors can establish with students a set of shared values or principles (e.g. fairness, equal access to opportunity, accessible technology). With these values in mind, and once students are engaged with the `hook,' then the instructor can more gently introduce the students to topics such as systematic racism that students might not be as open to if the instructor were to start directly at systematic racism.

\subsubsection{Finding a Responsible Computing Community of Collaborators}
\label{sec:rcc}
It is somewhat unlikely that a traditionally trained computing instructor by themselves would have all the knowledge to create a responsible computing intervention (unlike say a lesson in a more traditional computing course) mainly because the traditional training of computing students did not include much discussion of responsible computing. Here we consider two groups: researchers and faculty members from the humanities and the (target) students.

\paragraph{Working across disciplines}
As we have alluded to earlier, the humanities and social sciences have been studying how people form groups, how those groups interact with each other, and how those groups form societies. Within those societies, people produce texts, artifacts, and technologies that are shaped by and that continue to shape the values held by those societies. Hence, researchers and faculty members in these areas are natural domain experts on societal aspects of computing.

We recommend that if the instructor is thinking of creating a responsible computing intervention that they include an expert from the humanities and/or social sciences as a co-creator of the responsible computing intervention. While it is certainly possible to create an individual assignment/problem without (much) input from outside experts, we recommend that  the other four interventions be co-created with an expert in humanities and/or social sciences.

However, working across disciplines is hard. Specifically, building a collaboration with experts in other disciplines needs time and the instructor should have the time and willingness to put in effort to build such a cross-disciplinary collaboration. One important thing to keep in mind is to make sure that the collaboration is on an equal footing: specifically computing as a discipline tends exist with a  power of position over other disciplines especially those in the humanities and any such collaboration should be cognizant of such power differences and should consciously address these imbalances.

\paragraph{Students as Collaborators}

Ultimately, the success of a responsible computing intervention rests with how well the students respond to a specific intervention. Hence, it is vitally important that students have input in the creation process. We recommend that students from the target population (e.g. Teaching Assistants (TAs)-- even at the UG level) actually be involved in the design/decision making process during the creation of the intervention. For more details, see the Mozilla playbook section on student team dynamics~\cite{rcsplaybookstudentdynamics}.

A related concern is the availability of TAs who are familiar enough with the background material to be effective TAs for responsible computing intervention. This is not a concern for the intervention with individual assignments/problem (since at the end of the day those still work with traditional computing material) but are a concern for the other interventions especially when the intervention is offered for the first time (since there are no students who have taken the intervention in the previous offering of the course to recruit as TAs). One option is to recruit students from the humanities and/or social sciences. For more details, see the Mozilla playbook section on managing a teaching team~\cite{rcsplaybookteachingteam}.

\subsubsection{Responsible Computing and Social Justice}
\label{sec:sj}
While Responsible Computing is related to Social Justice there are some fundamental ways in which they differ. While the difference warrants a much longer and nuanced discussion, due to lack of space, we present an \textit{overly simplified} comparison of the two:
\footnote{We thank Kimberly Boulden and Dalia Muller for helping us articulate these differences.}
\begin{itemize}
    \item Computing Ethics is typically concerned with how to decide what is right vs. what is wrong while Social Justice asks what might be right vs. wrong \textit{for whom};
    \item Responsible Computing is concerned about doing the right thing while Social Justice asks \textit{who gets to decide} what is the right or wrong thing to do;
    \item Computing ethics typically can be more abstract and idealized in nature (often using predefined case studies as mentioned previously) while Social Justice is more about putting thought \textit{into practice} and in a current social context.
\end{itemize}

Of course the above comparison is not meant to pass judgment on which of two is `better'-- indeed we need both in our interventions. However, ethics/responsible computing tends to lend itself better to traditional computational ways of thinking where we are looking to define what we mean to be ethical/responsible and then to operationalize them using traditional computational/optimization techniques, such as logic. Or put another way, ethics/responsible computing ways of thinking can seem more in line with the traditional computing training where someone gives us the question that needs to be solved and we then subdivide the questions into smaller problems until we can solve the small sub-problems and piece them together to solve the original problem. However, traditionally computing takes the question as a given starting point, free from identified assumptions and often removed from a historical context. Social justice on the other hand encourages us to \textit{question the question itself}. Specifically, given how society is built today (with all the structural issues at play), social justice encourages us to ask whether the question being asked is the right question to ask– i.e. instead of breaking the question down into smaller problems the task is to `go back up' and reason about the question at the societal level, and ask what assumptions are informing how we understand the problem or situation in the first place

As alluded to above, social justice in some sense has a very different epistemology (way of knowing) and goals than the traditional computing training. This implies that computing students (especially those who have not thought a lot about questions that are conventional in the humanities) will have a hard time trying to `reset' their thinking. We recommend that the instructors carve out space in their course to give enough time for computing students to recognize how a computing epistemology frames problems and solutions in particular ways in order to help students get used to thinking in terms of social justice. A discussion of how a seemingly neutral rule can result in disparate impact via algorithmically driven decision systems is a great way to explore these ideas. This could e.g. mean that there could be dedicated course time (either via lectures or better yet discussions) to cover social justice related material (it would be more beneficial if this material was firmly rooted in computing applications). For details see the Mozilla playbook section on discussing justice and equity~\cite{rcsplaybookjustice}.

\paragraph{Community Involvement and Outreach}
One great way to motivate the students to work through the responsible computing intervention is to situate the intervention within the local community. For example, a local community organization would be the `client' of a project that students work on. The advantages for the students are apparent: they potentially get to make a difference in their local community, which can be a powerful motivating factor for the students.

However, involving the local community needs a lot of work before the course starts: specifically it needs to start with building a relationship between the instructor and the local organization. Special care must be taken that the relationship is mutually respectful by having the organization define its own needs and goals for the collaboration and e.g. the instructor must not have the mindset that they are doing the organization “a favor.” For more details, see the Mozilla playbook section on service learning~\cite{rcsplaybookservicelearning}.

\subsubsection{Learning Objectives and Assessment}
\label{sec:LO}
While learning outcomes and assessments are important for any course/lesson plan, for traditional computing courses/topics at some level the learning outcomes are well understood and the course assignments are set up to assess whether students achieve those outcomes or not. On the other hand, if the instructor is introducing a new responsible computing intervention in their course, it is even more important to both:

\begin{itemize}
    \item Clearly state the learning objectives of the intervention (which in turn would help in the actual design of the intervention); and
    \item Assess whether students achieve those learning objectives via explicitly setting up an assessment plan.
\end{itemize}

In addition to using existing student submissions for the interventions (e.g. reflection questions might be better suited to assess learning outcomes related to responsible computing), the instructor can also design specific assessment tools. For example, an instructor could use pre- and post-surveys to get a sense of how students are thinking about some higher level learning objectives that might not be that easy to glean by analyzing the student submissions. For more details, see the Mozilla playbook section on learning outcomes and assessment~\cite{rcsplaybooklearningoutcome}. They might also consult the SEP section of the ACM/IEEE 2023 curriculum guide.

We recommend that the instructors design the learning objectives along with an assessment plan to see if the intervention does actually help students achieve the learning objectives.

\subsubsection{Assessment}
\label{sec:assess}
Traditional computing courses tend to have assignments/projects that use traditional instruments to assess students: programming or proof based tasks. While these are still valuable instruments to be used in responsible computing intervention (especially since that allows students to see the tools that they have seen in other traditional computing courses in the context of responsible computing), such instruments by themselves are not enough. In our experience, we have noticed that even in an individual assignments/problem intervention if all the students have to submit are code (that is tested for traditional metrics like efficiency) or proofs (that is tested for traditional metrics like correctness) then the students will focus on those aspects of the intervention and not pay attention to the responsible computing aspects of the intervention. This in turn defeats the goal of the whole intervention.

We recommend that such traditional computing instruments be supplemented with more humanistic instruments e.g. reflection pieces. More specifically, each traditional instrument (e.g. writing a piece of code) should be accompanied by asking students to write (short) essay type answers where they reflect on the design choices they made in their code specifically with respect to their societal implications.

\subsection{What to do during the lesson/course?}
\label{sec:during}

In this section, we present some questions that the instructor should consider during the course. While most of the questions we recommend asking before the course starts are also relevant during the course, here we present questions that we have not considered so far. Most of these questions should be considered in parallel instead of in sequence (hence we recommend not to read too much into the ordering). \Cref{tab:during-class} provides our recommendations of which questions to consider when planning for specific interventions.

\subsubsection{Difficult Conversations}
\label{sec:difficult}
If the responsible computing intervention involves discussion of difficult topics (this would almost certainly be true for dedicated courses but might also be relevant for the other interventions), then the instructor needs to make sure that these discussions are set up in a way that leads to productive (though most like uncomfortable) discussions. We also note that some of the work of ensuring such productive discussions would need to be done before the course starts (e.g. what should the instructor do if a student cannot participate in a conversation because they might not be in the proper mental space to do so). Please see the Mozilla playbook section on having difficult conversations for more questions to consider~\cite{rcsplaybookdifficultconversations}. In addition, we recommend that the instructor consider the following additional points:

\begin{itemize}
    \item How to allow students to express themselves but have guard rails in place to make sure the conversation stays on topic;
    \item To make sure the students realize that these discussions are not a debate where “one has to win.” In addition to stating this point to the students, it should be reinforced in how students are graded in the discussions and how the instructor moderates the discussion.
    \item Consider allowing students to decide on the norms of discussion (if it makes sense,  the instructor can seed in some of the norms that are needed but students could as a whole pick a few more norms).

\end{itemize}

\subsubsection{Student Motivation}
\label{sec:student-motivate}
We have already coveted student apathy earlier in this section. However, making sure students are motivated about the topic is something that the instructor should pay attention to throughout the duration of the course. Note that motivation is not just picking a proper hook so that students can stay engaged but also challenging students to unlearn things they might take as a given. For example, insisting technology is neutral or more pertinently “staying neutral” is also a choice. Another pertinent example is to educate students about unanticipated consequences (see the Mozilla playbook section on unanticipated consequences~\cite{rcsplaybookunanticipated}).
Another common response of students to thinking about responsible computing is the refrain that if their future employers do not care about ethics/responsible computing then they will not be able to utilize what they learned (and hence what is the use of learning this material?). This points to the question of the agency of students to affect change is an important component that should be addressed. We recommend that (esp. in dedicated courses) responsible computing interventions talk about movement building among tech workers \cite{costanza2020design} For other ideas (including some outside the classroom), see the playbook section on Conversations about Responsible Computing and Employment Choices~\cite{rcsplaybookconversations}.

\subsubsection{Assessing Learning}
\label{sec:assess-learning}
We have already talked about assessments so we will keep this brief: (especially in dedicated courses) one should be open to changing plans  if the students are not responding in the way you would expect (doing some assessment during the semester might help or such a signal can also be gleaned from student submissions). For example, perhaps the instructor realizes that students need even more background in thinking about societal structures and it would make the rest of the course more fruitful if more time is given to students to think about such issues.

\subsection{What to do after the lesson/course?}
\label{sec:after}

Once the instructor has finished running the course (and the grades etc. have been submitted) it’s time to reflect on what went right and what did not. This exercise is useful at the end of a traditional computing course as well but here are some things that might be more specific to a responsible computing interventions:

\begin{itemize}
    \item Analyze the intervention specific assessment that was designed for the course and figure out how the course can be improved next time. Perhaps it would be better to co-teach the course with a colleague from the humanities and/or the social sciences?
    \item Students who really enjoyed the course will be looking for more to do in the future. If possible, be ready to connect them to the “next steps.” The Mozilla playbook section on making lessons stick has some starting points~\cite{rcsplaybooklessonsstick}.
\end{itemize}

\section{Examples for Teaching Society, Ethics, and Professionalism (SEP) Topics}
\label{sec:examples}

\Cref{app:examples} (Tables~\ref{tab:indv-prob-ex}-\ref{tab:dedicated-ex}) highlight illustrative examples of teaching SEP topics as part of a curriculum in a traditional computer computing course and SEP taught as a dedicated course. There is no one size fits all solution for incorporating SEP into computing programs; however these examples are intended to highlight a range of approaches to support adaptation in different academic environments. Links to the resources and the larger body of work are provided. Other resources can be found on the \hyperlink{https://www.engage-csedu.org}{ACM EngageCSEdu site }which is starting to solicit instructional materials with embedded responsible computing content.

\section{Conclusions}
\label{sec:concl}

The goal of this paper was to provide a foundational rationale for why and how to embed responsible computing principles, practices, and lessons into the undergraduate CS curriculum. It also provides practical advice on a number of different approaches with examples of other instructors who are engaged in the work. The growing interest and community in the ethical responsibilities of computing professionals means that there are many resources available for review and adoption as well as groups of faculty who have found creative ways to engage students in discussions, activities, and projects focused on both computing concepts and skills as well as ethical practices. In the appendices, we provide illustrative examples with links to open source curriculum resources. By sharing high quality lessons, we hope to encourage instructors who may be interested but intimidated or reluctant to start small, choosing one topic to try using some of the suggested approaches and content. We also hope to make connecting the new SEP areas to existing curriculum easier during the transition to the new CS2023 curriculum.


\begin{acks}
This work was done as part of ACM/IEEE-CS/AAAI Computer Science Curricula 2023 (\href{https://csed.acm.org}{csed.acm.org}).

We would like to acknowledge and thank all of the responsible computing educators and advocates who have been doing this work for a long time. 
\end{acks}

\bibliographystyle{ACM-Reference-Format}
\bibliography{citations}


\begin{thebibliography}{62}


\ifx \showCODEN    \undefined \def \showCODEN     #1{\unskip}     \fi
\ifx \showDOI      \undefined \def \showDOI       #1{#1}\fi
\ifx \showISBNx    \undefined \def \showISBNx     #1{\unskip}     \fi
\ifx \showISBNxiii \undefined \def \showISBNxiii  #1{\unskip}     \fi
\ifx \showISSN     \undefined \def \showISSN      #1{\unskip}     \fi
\ifx \showLCCN     \undefined \def \showLCCN      #1{\unskip}     \fi
\ifx \shownote     \undefined \def \shownote      #1{#1}          \fi
\ifx \showarticletitle \undefined \def \showarticletitle #1{#1}   \fi
\ifx \showURL      \undefined \def \showURL       {\relax}        \fi
\providecommand\bibfield[2]{#2}
\providecommand\bibinfo[2]{#2}
\providecommand\natexlab[1]{#1}
\providecommand\showeprint[2][]{arXiv:#2}

\bibitem[bem(2022)]%
        {bemidjirepo}
 \bibinfo{year}{"2022"}\natexlab{}.
\newblock \bibinfo{booktitle}{\emph{"Responsible Computer Science through
  Engaged Collaboration"}}.
\newblock
\urldef\tempurl%
\url{"https://www.bemidjistate.edu/academics/departments/mathematics-computer-science/rcs/"}
\showURL{%
\tempurl}


\bibitem[eng(2023)]%
        {engagecseduethicsrepo}
 \bibinfo{year}{"2023"}\natexlab{}.
\newblock \bibinfo{booktitle}{\emph{"ACM EngageCSedu Ethics Repository"}}.
\newblock
\urldef\tempurl%
\url{"https://www.engage-csedu.org/ethics-and-computing/repository"}
\showURL{%
\tempurl}


\bibitem[com(2023)]%
        {computingnarrativesrepo}
 \bibinfo{year}{"2023"}\natexlab{}.
\newblock \bibinfo{booktitle}{\emph{"Computing Ethics Narratives Repository at
  Bowdoin College and Collby College}}.
\newblock
\urldef\tempurl%
\url{"https://computingnarratives.com"}
\showURL{%
\tempurl}


\bibitem[emb(2023)]%
        {embedethicsharvardrepo}
 \bibinfo{year}{"2023"}\natexlab{}.
\newblock \bibinfo{booktitle}{\emph{"Embedded EthiCS @ Harvard University -
  Modules Repository"}}.
\newblock


\bibitem[eth(2023)]%
        {ethics4eu}
 \bibinfo{year}{"2023"}\natexlab{}.
\newblock \bibinfo{booktitle}{\emph{Ethics 4 EU - Educational Resources"}}.
\newblock
\urldef\tempurl%
\url{"https://ascnet.ie/ethics4eu-website/welcome-to-the-bricks/"}
\showURL{%
\tempurl}


\bibitem[pec(2023)]%
        {peckethics}
 \bibinfo{year}{2023}\natexlab{}.
\newblock \bibinfo{booktitle}{\emph{Integrating Social Responsibility into Core
  CS}}.
\newblock
\urldef\tempurl%
\url{https://evanpeck.github.io/projects/responsibleCS}
\showURL{%
\tempurl}


\bibitem[ABET(2023)]%
        {abet-cac}
\bibfield{author}{\bibinfo{person}{ABET}.} \bibinfo{year}{2023}\natexlab{}.
\newblock \bibinfo{booktitle}{\emph{Criteria for Accrediting Computing
  Programs, 2023 – 2024}}.
\newblock
\urldef\tempurl%
\url{https://www.abet.org/accreditation/accreditation-criteria/criteria-for-accrediting-computing-programs-2023-2024/}
\showURL{%
\tempurl}
\newblock
\shownote{Last Accessed: November 24, 2023}.


\bibitem[Benjamin(2023)]%
        {benjamin2023race}
\bibfield{author}{\bibinfo{person}{Ruha Benjamin}.}
  \bibinfo{year}{2023}\natexlab{}.
\newblock \showarticletitle{Race after technology}.
\newblock In \bibinfo{booktitle}{\emph{Social Theory Re-Wired}}.
  \bibinfo{publisher}{Routledge}, \bibinfo{pages}{405--415}.
\newblock


\bibitem[Blumenthal(2021)]%
        {blumenthal2021walking}
\bibfield{author}{\bibinfo{person}{Richard Blumenthal}.}
  \bibinfo{year}{2021}\natexlab{}.
\newblock \showarticletitle{Walking the curricular talk: A longitudinal study
  of computer science departmental course requirements}.
\newblock \bibinfo{journal}{\emph{Journal of Computing Sciences in Colleges}}
  \bibinfo{volume}{37}, \bibinfo{number}{2} (\bibinfo{year}{2021}),
  \bibinfo{pages}{40--50}.
\newblock


\bibitem[Boenig-Liptsin(2020)]%
        {rcsplaybooklearningoutcome}
\bibfield{author}{\bibinfo{person}{Margo Boenig-Liptsin}.}
  \bibinfo{year}{2020}\natexlab{}.
\newblock \bibinfo{booktitle}{\emph{Learning Outcomes and Assessments}}.
\newblock In Pham and Rudra \citeN{rcscplaybookoverview2020}.
\newblock
\urldef\tempurl%
\url{https://foundation.mozilla.org/en/what-we-fund/awards/teaching-responsible-computing-playbook/topics/learning-outcomes-and-assessments/}
\showURL{%
\tempurl}
\newblock
\shownote{Last Accessed Nov 11, 2023.}.


\bibitem[Boenig-Liptsin and Carson(2020)]%
        {rcsplaybookconversations}
\bibfield{author}{\bibinfo{person}{Margo Boenig-Liptsin} {and}
  \bibinfo{person}{Cathryn Carson}.} \bibinfo{year}{2020}\natexlab{}.
\newblock \bibinfo{booktitle}{\emph{Conversations about Responsible Computing
  and Employment Choices}}.
\newblock In Pham and Rudra \citeN{rcscplaybookoverview2020}.
\newblock
\urldef\tempurl%
\url{https://foundation.mozilla.org/en/what-we-fund/awards/teaching-responsible-computing-playbook/topics/employment-choices/}
\showURL{%
\tempurl}
\newblock
\shownote{Last Accessed Nov 11, 2023.}.


\bibitem[Boenig-Liptsin and Liu(2020)]%
        {rcsplaybookstudentdynamics}
\bibfield{author}{\bibinfo{person}{Margo Boenig-Liptsin} {and}
  \bibinfo{person}{Xin Liu}.} \bibinfo{year}{2020}\natexlab{}.
\newblock \bibinfo{booktitle}{\emph{Student Team Dynamics}}.
\newblock In Pham and Rudra \citeN{rcscplaybookoverview2020}.
\newblock
\urldef\tempurl%
\url{https://foundation.mozilla.org/en/what-we-fund/awards/teaching-responsible-computing-playbook/topics/student-team-dynamics/}
\showURL{%
\tempurl}
\newblock
\shownote{Last Accessed Nov 11, 2023.}.


\bibitem[Boenig-Liptsin and Ricks(2020)]%
        {rcsplaybookjustice}
\bibfield{author}{\bibinfo{person}{Margo Boenig-Liptsin} {and}
  \bibinfo{person}{Vance Ricks}.} \bibinfo{year}{2020}\natexlab{}.
\newblock \bibinfo{booktitle}{\emph{Discussing Justice and Equity}}.
\newblock In Pham and Rudra \citeN{rcscplaybookoverview2020}.
\newblock
\urldef\tempurl%
\url{https://foundation.mozilla.org/en/what-we-fund/awards/teaching-responsible-computing-playbook/topics/discuss-justice-equity/}
\showURL{%
\tempurl}
\newblock
\shownote{Last Accessed Nov 11, 2023.}.


\bibitem[Bonham-Carter et~al\mbox{.}(2020)]%
        {rcsplaybookservicelearning}
\bibfield{author}{\bibinfo{person}{Oliver Bonham-Carter},
  \bibinfo{person}{Antonio Delgado}, \bibinfo{person}{George Gabb}, {and}
  \bibinfo{person}{Joshua Young}.} \bibinfo{year}{2020}\natexlab{}.
\newblock \bibinfo{booktitle}{\emph{Service Learning}}.
\newblock In Pham and Rudra \citeN{rcscplaybookoverview2020}.
\newblock
\urldef\tempurl%
\url{https://foundation.mozilla.org/en/what-we-fund/awards/teaching-responsible-computing-playbook/topics/service-learning/}
\showURL{%
\tempurl}
\newblock
\shownote{Last Accessed Nov 11, 2023.}.


\bibitem[Brey(2000a)]%
        {brey2000disclosive}
\bibfield{author}{\bibinfo{person}{Philip Brey}.}
  \bibinfo{year}{2000}\natexlab{a}.
\newblock \showarticletitle{Disclosive computer ethics}.
\newblock \bibinfo{journal}{\emph{ACM Sigcas Computers and Society}}
  \bibinfo{volume}{30}, \bibinfo{number}{4} (\bibinfo{year}{2000}),
  \bibinfo{pages}{10--16}.
\newblock


\bibitem[Brey(2000b)]%
        {brey2000method}
\bibfield{author}{\bibinfo{person}{Philip Brey}.}
  \bibinfo{year}{2000}\natexlab{b}.
\newblock \showarticletitle{Method in computer ethics: Towards a multi-level
  interdisciplinary approach}.
\newblock \bibinfo{journal}{\emph{Ethics and information technology}}
  \bibinfo{volume}{2}, \bibinfo{number}{2} (\bibinfo{year}{2000}),
  \bibinfo{pages}{125--129}.
\newblock


\bibitem[Buolamwini and Gebru(2018)]%
        {buolamwini2018gender}
\bibfield{author}{\bibinfo{person}{Joy Buolamwini} {and}
  \bibinfo{person}{Timnit Gebru}.} \bibinfo{year}{2018}\natexlab{}.
\newblock \showarticletitle{Gender shades: Intersectional accuracy disparities
  in commercial gender classification}. In \bibinfo{booktitle}{\emph{Conference
  on fairness, accountability and transparency}}. "PMLR",
  \bibinfo{pages}{77--91}.
\newblock


\bibitem[Burton et~al\mbox{.}(2018)]%
        {burton2018teach}
\bibfield{author}{\bibinfo{person}{Emanuelle Burton}, \bibinfo{person}{Judy
  Goldsmith}, {and} \bibinfo{person}{Nicholas Mattei}.}
  \bibinfo{year}{2018}\natexlab{}.
\newblock \showarticletitle{How to teach computer ethics through science
  fiction}.
\newblock \bibinfo{journal}{\emph{Commun. ACM}} \bibinfo{volume}{61},
  \bibinfo{number}{8} (\bibinfo{year}{2018}), \bibinfo{pages}{54--64}.
\newblock


\bibitem[Burton et~al\mbox{.}(2023)]%
        {burton2023computing}
\bibfield{author}{\bibinfo{person}{Emanuelle Burton}, \bibinfo{person}{Judy
  Goldsmith}, \bibinfo{person}{Nicholas Mattei}, \bibinfo{person}{Cory Siler},
  {and} \bibinfo{person}{Sara-Jo Swiatek}.} \bibinfo{year}{2023}\natexlab{}.
\newblock \bibinfo{booktitle}{\emph{Computing and Technology Ethics: Engaging
  through Science Fiction}}.
\newblock \bibinfo{publisher}{MIT Press}.
\newblock


\bibitem[Califf and Goodwin(2005)]%
        {califf2005effective}
\bibfield{author}{\bibinfo{person}{Mary~Elaine Califf} {and}
  \bibinfo{person}{Mary Goodwin}.} \bibinfo{year}{2005}\natexlab{}.
\newblock \showarticletitle{Effective incorporation of ethics into courses that
  focus on programming}.
\newblock \bibinfo{journal}{\emph{ACM SIGCSE Bulletin}} \bibinfo{volume}{37},
  \bibinfo{number}{1} (\bibinfo{year}{2005}), \bibinfo{pages}{347--351}.
\newblock


\bibitem[Cech(2014)]%
        {cech2014culture}
\bibfield{author}{\bibinfo{person}{Erin~A Cech}.}
  \bibinfo{year}{2014}\natexlab{}.
\newblock \showarticletitle{Culture of disengagement in engineering education?}
\newblock \bibinfo{journal}{\emph{Science, Technology, \& Human Values}}
  \bibinfo{volume}{39}, \bibinfo{number}{1} (\bibinfo{year}{2014}),
  \bibinfo{pages}{42--72}.
\newblock


\bibitem[Cohen et~al\mbox{.}(2021)]%
        {cohen2021new}
\bibfield{author}{\bibinfo{person}{Lena Cohen}, \bibinfo{person}{Heila Precel},
  \bibinfo{person}{Harold Triedman}, {and} \bibinfo{person}{Kathi Fisler}.}
  \bibinfo{year}{2021}\natexlab{}.
\newblock \showarticletitle{A New Model for Weaving Responsible Computing Into
  Courses Across the CS Curriculum}. In \bibinfo{booktitle}{\emph{Proceedings
  of the 52nd ACM Technical Symposium on Computer Science Education}}.
  \bibinfo{pages}{858--864}.
\newblock


\bibitem[Costanza-Chock(2020)]%
        {costanza2020design}
\bibfield{author}{\bibinfo{person}{Sasha Costanza-Chock}.}
  \bibinfo{year}{2020}\natexlab{}.
\newblock \bibinfo{booktitle}{\emph{Design justice: Community-led practices to
  build the worlds we need}}.
\newblock \bibinfo{publisher}{The MIT Press}.
\newblock


\bibitem[Cytron and Das(2020)]%
        {rcsplaybookteachingteam}
\bibfield{author}{\bibinfo{person}{Ron Cytron} {and} \bibinfo{person}{Udayan
  Das}.} \bibinfo{year}{2020}\natexlab{}.
\newblock \bibinfo{booktitle}{\emph{Manage Teaching Team}}.
\newblock In Pham and Rudra \citeN{rcscplaybookoverview2020}.
\newblock
\urldef\tempurl%
\url{https://foundation.mozilla.org/en/what-we-fund/awards/teaching-responsible-computing-playbook/topics/manage-teaching-team/}
\showURL{%
\tempurl}
\newblock
\shownote{Last Accessed Nov 11, 2023.}.


\bibitem[Davis and Walker(2011)]%
        {davis2011incorporating}
\bibfield{author}{\bibinfo{person}{Janet Davis} {and} \bibinfo{person}{Henry~M
  Walker}.} \bibinfo{year}{2011}\natexlab{}.
\newblock \showarticletitle{Incorporating social issues of computing in a
  small, liberal arts college: a case study}. In
  \bibinfo{booktitle}{\emph{Proceedings of the 42nd ACM technical symposium on
  Computer science education}}. \bibinfo{pages}{69--74}.
\newblock


\bibitem[Doore and Yunus(2022)]%
        {doore2022release}
\bibfield{author}{\bibinfo{person}{Stacy~A Doore} {and} \bibinfo{person}{Azalea
  Yunus}.} \bibinfo{year}{2022}\natexlab{}.
\newblock \showarticletitle{Release the robot dogs}.
\newblock \bibinfo{journal}{\emph{ETHICOMP 2022}} (\bibinfo{year}{2022}),
  \bibinfo{pages}{188}.
\newblock


\bibitem[Ferreira and Vardi(2020)]%
        {ferreira2020computer}
\bibfield{author}{\bibinfo{person}{Rodrigo Ferreira} {and}
  \bibinfo{person}{Moshe~Y Vardi}.} \bibinfo{year}{2020}\natexlab{}.
\newblock \showarticletitle{Computer Ethics and Care}.
\newblock  (\bibinfo{year}{2020}).
\newblock


\bibitem[Fiesler et~al\mbox{.}(2021)]%
        {fiesler2021integrating}
\bibfield{author}{\bibinfo{person}{Casey Fiesler}, \bibinfo{person}{Mikhaila
  Friske}, \bibinfo{person}{Natalie Garrett}, \bibinfo{person}{Felix Muzny},
  \bibinfo{person}{Jessie~J Smith}, {and} \bibinfo{person}{Jason Zietz}.}
  \bibinfo{year}{2021}\natexlab{}.
\newblock \showarticletitle{Integrating Ethics into Introductory Programming
  Classes}. In \bibinfo{booktitle}{\emph{Proceedings of the 52nd ACM Technical
  Symposium on Computer Science Education}}. \bibinfo{pages}{1027--1033}.
\newblock


\bibitem[Fiesler et~al\mbox{.}(2020)]%
        {fiesler2020we}
\bibfield{author}{\bibinfo{person}{Casey Fiesler}, \bibinfo{person}{Natalie
  Garrett}, {and} \bibinfo{person}{Nathan Beard}.}
  \bibinfo{year}{2020}\natexlab{}.
\newblock \showarticletitle{What do we teach when we teach tech ethics? A
  syllabi analysis}. In \bibinfo{booktitle}{\emph{Proceedings of the 51st ACM
  technical symposium on computer science education}}.
  \bibinfo{pages}{289--295}.
\newblock


\bibitem[Fisler et~al\mbox{.}(2022)]%
        {fisler2022approaches}
\bibfield{author}{\bibinfo{person}{Kathi Fisler}, \bibinfo{person}{Sorelle
  Friedler}, \bibinfo{person}{Kevin Lin}, {and} \bibinfo{person}{Suresh
  Venkatasubramanian}.} \bibinfo{year}{2022}\natexlab{}.
\newblock \showarticletitle{Approaches for Weaving Responsible Computing into
  Data Structures and Algorithms Courses}. In
  \bibinfo{booktitle}{\emph{Proceedings of the 53rd ACM Technical Symposium on
  Computer Science Education V. 2}}. \bibinfo{pages}{1049--1050}.
\newblock


\bibitem[Garrett et~al\mbox{.}(2020)]%
        {garrett2020more}
\bibfield{author}{\bibinfo{person}{Natalie Garrett}, \bibinfo{person}{Nathan
  Beard}, {and} \bibinfo{person}{Casey Fiesler}.}
  \bibinfo{year}{2020}\natexlab{}.
\newblock \showarticletitle{More than" If Time Allows" the role of ethics in AI
  education}. In \bibinfo{booktitle}{\emph{Proceedings of the AAAI/ACM
  Conference on AI, Ethics, and Society}}. \bibinfo{pages}{272--278}.
\newblock


\bibitem[Goetze(2023)]%
        {goetz2023}
\bibfield{author}{\bibinfo{person}{Trystan~S. Goetze}.}
  \bibinfo{year}{2023}\natexlab{}.
\newblock \showarticletitle{Integrating Ethics into Computer Science Education:
  Multi-, Inter-, and Transdisciplinary Approaches}.
  \bibinfo{publisher}{Association for Computing Machinery},
  \bibinfo{address}{New York, NY, USA}.
\newblock
\showISBNx{9781450394314}
\urldef\tempurl%
\url{https://doi.org/10.1145/3545945.3569792}
\showDOI{\tempurl}


\bibitem[Goldweber et~al\mbox{.}(2022)]%
        {goldweber2022hands}
\bibfield{author}{\bibinfo{person}{Mikey Goldweber}, \bibinfo{person}{Lisa
  Kaczmarczyk}, \bibinfo{person}{Rick Blumenthal}, \bibinfo{person}{Alison
  Clear}, {and} \bibinfo{person}{Johanna Blumenthal}.}
  \bibinfo{year}{2022}\natexlab{}.
\newblock \showarticletitle{A Hands-On Tutorial on How To Incorporate Computing
  for Social Good in the Introductory Course Sequence}. In
  \bibinfo{booktitle}{\emph{Proceedings of the 53rd ACM Technical Symposium on
  Computer Science Education V. 2}}. \bibinfo{pages}{1039--1040}.
\newblock


\bibitem[Greer and Wolf(2020)]%
        {greer2020overcoming}
\bibfield{author}{\bibinfo{person}{Colleen Greer} {and}
  \bibinfo{person}{Marty~J Wolf}.} \bibinfo{year}{2020}\natexlab{}.
\newblock \showarticletitle{Overcoming barriers to including ethics and social
  responsibility in computing courses}. In \bibinfo{booktitle}{\emph{Societal
  Challenges in the Smart Society}}. Universidad de La Rioja,
  \bibinfo{pages}{131--144}.
\newblock


\bibitem[Grosz et~al\mbox{.}(2019)]%
        {grosz2019embedded}
\bibfield{author}{\bibinfo{person}{Barbara~J Grosz},
  \bibinfo{person}{David~Gray Grant}, \bibinfo{person}{Kate Vredenburgh},
  \bibinfo{person}{Jeff Behrends}, \bibinfo{person}{Lily Hu},
  \bibinfo{person}{Alison Simmons}, {and} \bibinfo{person}{Jim Waldo}.}
  \bibinfo{year}{2019}\natexlab{}.
\newblock \showarticletitle{Embedded EthiCS: integrating ethics across CS
  education}.
\newblock \bibinfo{journal}{\emph{Commun. ACM}} \bibinfo{volume}{62},
  \bibinfo{number}{8} (\bibinfo{year}{2019}), \bibinfo{pages}{54--61}.
\newblock


\bibitem[Horton et~al\mbox{.}(2023)]%
        {horton2023more}
\bibfield{author}{\bibinfo{person}{Diane Horton}, \bibinfo{person}{David Liu},
  \bibinfo{person}{Sheila~A McIlraith}, {and} \bibinfo{person}{Nina Wang}.}
  \bibinfo{year}{2023}\natexlab{}.
\newblock \showarticletitle{Is More Better When Embedding Ethics in CS
  Courses?}. In \bibinfo{booktitle}{\emph{Proceedings of the 54th ACM Technical
  Symposium on Computer Science Education V. 1}}. \bibinfo{pages}{652--658}.
\newblock


\bibitem[Horton et~al\mbox{.}(2022)]%
        {horton2022embedding}
\bibfield{author}{\bibinfo{person}{Diane Horton}, \bibinfo{person}{Sheila~A
  McIlraith}, \bibinfo{person}{Nina Wang}, \bibinfo{person}{Maryam Majedi},
  \bibinfo{person}{Emma McClure}, {and} \bibinfo{person}{Benjamin Wald}.}
  \bibinfo{year}{2022}\natexlab{}.
\newblock \showarticletitle{Embedding Ethics in Computer Science Courses: Does
  it Work?}. In \bibinfo{booktitle}{\emph{Proceedings of the 53rd ACM Technical
  Symposium on Computer Science Education V. 1}}. \bibinfo{pages}{481--487}.
\newblock


\bibitem[Johnson(1994)]%
        {johnson1994computer}
\bibfield{author}{\bibinfo{person}{Deborah~G Johnson}.}
  \bibinfo{year}{1994}\natexlab{}.
\newblock \bibinfo{booktitle}{\emph{Computer ethics}}.
\newblock \bibinfo{publisher}{Prentice-Hall, Inc.}
\newblock


\bibitem[Kamara(2020)]%
        {rcsplaybookdifficultconversations}
\bibfield{author}{\bibinfo{person}{Seny Kamara}.}
  \bibinfo{year}{2020}\natexlab{}.
\newblock \bibinfo{booktitle}{\emph{Difficult Conversations}}.
\newblock In Pham and Rudra \citeN{rcscplaybookoverview2020}.
\newblock
\urldef\tempurl%
\url{https://foundation.mozilla.org/en/what-we-fund/awards/teaching-responsible-computing-playbook/topics/difficult-conversations/}
\showURL{%
\tempurl}
\newblock
\shownote{Last Accessed Nov 11, 2023.}.


\bibitem[McDonald et~al\mbox{.}(2022)]%
        {mcdonald2022responsible}
\bibfield{author}{\bibinfo{person}{Nora McDonald}, \bibinfo{person}{Adegboyega
  Akinsiku}, \bibinfo{person}{Jonathan Hunter-Cevera}, \bibinfo{person}{Maria
  Sanchez}, \bibinfo{person}{Kerrie Kephart}, \bibinfo{person}{Mark
  Berczynski}, {and} \bibinfo{person}{Helena~M Mentis}.}
  \bibinfo{year}{2022}\natexlab{}.
\newblock \showarticletitle{Responsible Computing: A Longitudinal Study of a
  Peer-led Ethics Learning Framework}.
\newblock \bibinfo{journal}{\emph{ACM Transactions on Computing Education
  (TOCE)}} \bibinfo{volume}{22}, \bibinfo{number}{4} (\bibinfo{year}{2022}),
  \bibinfo{pages}{1--21}.
\newblock


\bibitem[Mehdiabadi(2019)]%
        {mehdiabadi2019facing}
\bibfield{author}{\bibinfo{person}{Amir~Hedayati Mehdiabadi}.}
  \bibinfo{year}{2019}\natexlab{}.
\newblock \showarticletitle{Facing Computer Ethics Dilemmas: Comparing Ethical
  Decision-Making Processes of Students in Computer Science with Non-Computer
  Science Majors}. In \bibinfo{booktitle}{\emph{2019 ASEE Annual Conference \&
  Exposition}}.
\newblock


\bibitem[Mentis and Ricks(2020)]%
        {rcsplaybooklessonsstick}
\bibfield{author}{\bibinfo{person}{Helena Mentis} {and} \bibinfo{person}{Vance
  Ricks}.} \bibinfo{year}{2020}\natexlab{}.
\newblock \bibinfo{booktitle}{\emph{Making Lessons Stick}}.
\newblock In Pham and Rudra \citeN{rcscplaybookoverview2020}.
\newblock
\urldef\tempurl%
\url{https://foundation.mozilla.org/en/what-we-fund/awards/teaching-responsible-computing-playbook/topics/lessons-stick/}
\showURL{%
\tempurl}
\newblock
\shownote{Last Accessed Nov 11, 2023.}.


\bibitem[Najibi(2020)]%
        {najibi2020racial}
\bibfield{author}{\bibinfo{person}{Alex Najibi}.}
  \bibinfo{year}{2020}\natexlab{}.
\newblock \showarticletitle{Racial discrimination in face recognition
  technology}.
\newblock \bibinfo{journal}{\emph{Science in the News}}  \bibinfo{volume}{24}
  (\bibinfo{year}{2020}).
\newblock


\bibitem[National Academies~of Sciences et~al\mbox{.}(2022)]%
        {national2022fostering}
\bibfield{author}{\bibinfo{person}{Engineering National Academies~of Sciences},
  \bibinfo{person}{Medicine}, {et~al\mbox{.}}} \bibinfo{year}{2022}\natexlab{}.
\newblock \showarticletitle{Fostering Responsible Computing Research:
  Foundations and Practices}.
\newblock  (\bibinfo{year}{2022}).
\newblock


\bibitem[Pham and Rudra(2020)]%
        {rcscplaybookoverview2020}
\bibfield{editor}{\bibinfo{person}{Kathy Pham} {and} \bibinfo{person}{Atri
  Rudra}} (Eds.). \bibinfo{year}{2020}\natexlab{}.
\newblock \bibinfo{booktitle}{\emph{"Teaching Responsible Computing Playbook}}.
\newblock
\urldef\tempurl%
\url{https://foundation.mozilla.org/en/what-we-fund/awards/teaching-responsible-computing-playbook/}
\showURL{%
\tempurl}
\newblock
\shownote{Last Accessed Nov 11, 2023.}.


\bibitem[Raj et~al\mbox{.}(2022)]%
        {raj2022interpreting}
\bibfield{author}{\bibinfo{person}{Rajendra~K Raj}, \bibinfo{person}{Amruth~N
  Kumar}, \bibinfo{person}{Mihaela Sabin}, {and} \bibinfo{person}{John
  Impagliazzo}.} \bibinfo{year}{2022}\natexlab{}.
\newblock \showarticletitle{Interpreting the ABET Computer Science Criteria
  Using Competencies}. In \bibinfo{booktitle}{\emph{Proceedings of the 53rd ACM
  Technical Symposium on Computer Science Education V. 1}}.
  \bibinfo{pages}{906--912}.
\newblock


\bibitem[Reich et~al\mbox{.}(2020)]%
        {reich2020teaching}
\bibfield{author}{\bibinfo{person}{Rob Reich}, \bibinfo{person}{Mehran Sahami},
  \bibinfo{person}{Jeremy~M Weinstein}, {and} \bibinfo{person}{Hilary Cohen}.}
  \bibinfo{year}{2020}\natexlab{}.
\newblock \showarticletitle{Teaching computer ethics: A deeply
  multidisciplinary approach}. In \bibinfo{booktitle}{\emph{Proceedings of the
  51st ACM Technical Symposium on Computer Science Education}}.
  \bibinfo{pages}{296--302}.
\newblock


\bibitem[Ricks(2020)]%
        {rcsplaybookunanticipated}
\bibfield{author}{\bibinfo{person}{Vance Ricks}.}
  \bibinfo{year}{2020}\natexlab{}.
\newblock \bibinfo{booktitle}{\emph{Talking About Unanticipated Consequences}}.
\newblock In Pham and Rudra \citeN{rcscplaybookoverview2020}.
\newblock
\urldef\tempurl%
\url{https://foundation.mozilla.org/en/what-we-fund/awards/teaching-responsible-computing-playbook/topics/unanticipated-consequences/}
\showURL{%
\tempurl}
\newblock
\shownote{Last Accessed Nov 11, 2023.}.


\bibitem[Rivero(2020)]%
        {rivero2020}
\bibfield{author}{\bibinfo{person}{Nicolas Rivero}.}
  \bibinfo{year}{2020}\natexlab{}.
\newblock \showarticletitle{The influential project that sparked the end of
  IBM’s facial recognition program}.
\newblock \bibinfo{journal}{\emph{Quartz}} (\bibinfo{year}{2020}).
\newblock
\urldef\tempurl%
\url{https://qz.com/1866848/why-ibm-abandoned-its-facial-recognition-program}
\showURL{%
\tempurl}


\bibitem[Rudra(2020)]%
        {rcsplaybookaccess}
\bibfield{author}{\bibinfo{person}{Atri Rudra}.}
  \bibinfo{year}{2020}\natexlab{}.
\newblock \bibinfo{booktitle}{\emph{Access to Technology}}.
\newblock In Pham and Rudra \citeN{rcscplaybookoverview2020}.
\newblock
\urldef\tempurl%
\url{https://foundation.mozilla.org/en/what-we-fund/awards/teaching-responsible-computing-playbook/topics/access-to-tech/}
\showURL{%
\tempurl}
\newblock
\shownote{Last Accessed Nov 11, 2023.}.


\bibitem[Sahami and Roach(2014)]%
        {sahami2014computer}
\bibfield{author}{\bibinfo{person}{Mehran Sahami} {and} \bibinfo{person}{Steve
  Roach}.} \bibinfo{year}{2014}\natexlab{}.
\newblock \showarticletitle{Computer science curricula 2013 released}.
\newblock \bibinfo{journal}{\emph{Commun. ACM}} \bibinfo{volume}{57},
  \bibinfo{number}{6} (\bibinfo{year}{2014}), \bibinfo{pages}{5--5}.
\newblock


\bibitem[Sanders(2005)]%
        {sanders2005discussion}
\bibfield{author}{\bibinfo{person}{Alton~F Sanders}.}
  \bibinfo{year}{2005}\natexlab{}.
\newblock \showarticletitle{A discussion format for computer ethics}. In
  \bibinfo{booktitle}{\emph{Proceedings of the 36th SIGCSE technical symposium
  on Computer science education}}. \bibinfo{pages}{352--355}.
\newblock


\bibitem[Shapiro et~al\mbox{.}(2021)]%
        {shapiro2021using}
\bibfield{author}{\bibinfo{person}{Ben~Rydal Shapiro}, \bibinfo{person}{Emma
  Lovegall}, \bibinfo{person}{Amanda Meng}, \bibinfo{person}{Jason Borenstein},
  {and} \bibinfo{person}{Ellen Zegura}.} \bibinfo{year}{2021}\natexlab{}.
\newblock \showarticletitle{Using Role-Play to Scale the Integration of Ethics
  Across the Computer Science Curriculum}. In
  \bibinfo{booktitle}{\emph{Proceedings of the 52nd ACM Technical Symposium on
  Computer Science Education}}. \bibinfo{pages}{1034--1040}.
\newblock


\bibitem[Sjoding et~al\mbox{.}(2020)]%
        {sjoding2020racial}
\bibfield{author}{\bibinfo{person}{Michael~W Sjoding},
  \bibinfo{person}{Robert~P Dickson}, \bibinfo{person}{Theodore~J Iwashyna},
  \bibinfo{person}{Steven~E Gay}, {and} \bibinfo{person}{Thomas~S Valley}.}
  \bibinfo{year}{2020}\natexlab{}.
\newblock \showarticletitle{Racial bias in pulse oximetry measurement}.
\newblock \bibinfo{journal}{\emph{New England Journal of Medicine}}
  \bibinfo{volume}{383}, \bibinfo{number}{25} (\bibinfo{year}{2020}),
  \bibinfo{pages}{2477--2478}.
\newblock


\bibitem[Skirpan et~al\mbox{.}(2018)]%
        {skirpan2018quantified}
\bibfield{author}{\bibinfo{person}{Michael Skirpan},
  \bibinfo{person}{Jacqueline Cameron}, {and} \bibinfo{person}{Tom Yeh}.}
  \bibinfo{year}{2018}\natexlab{}.
\newblock \showarticletitle{Quantified self: An interdisciplinary immersive
  theater project supporting a collaborative learning environment for cs
  ethics}. In \bibinfo{booktitle}{\emph{Proceedings of the 49th ACM Technical
  Symposium on Computer Science Education}}. \bibinfo{pages}{946--951}.
\newblock


\bibitem[Smith et~al\mbox{.}(2023)]%
        {smith2023incorporating}
\bibfield{author}{\bibinfo{person}{Jessie~J Smith}, \bibinfo{person}{Blakeley~H
  Payne}, \bibinfo{person}{Shamika Klassen}, \bibinfo{person}{Dylan~Thomas
  Doyle}, {and} \bibinfo{person}{Casey Fiesler}.}
  \bibinfo{year}{2023}\natexlab{}.
\newblock \showarticletitle{Incorporating Ethics in Computing Courses:
  Barriers, Support, and Perspectives from Educators}. In
  \bibinfo{booktitle}{\emph{Proceedings of the 54th ACM Technical Symposium on
  Computer Science Education V. 1}}. \bibinfo{pages}{367--373}.
\newblock


\bibitem[Smith et~al\mbox{.}(2021)]%
        {smith2021decolonizing}
\bibfield{author}{\bibinfo{person}{Rachel~Charlotte Smith},
  \bibinfo{person}{Heike Winschiers-Theophilus}, \bibinfo{person}{Daria Loi},
  \bibinfo{person}{Rog{\'e}rio~Abreu de Paula}, \bibinfo{person}{Asnath~Paula
  Kambunga}, \bibinfo{person}{Marly~Muudeni Samuel}, {and}
  \bibinfo{person}{Tariq Zaman}.} \bibinfo{year}{2021}\natexlab{}.
\newblock \showarticletitle{Decolonizing design practices: towards
  pluriversality}. In \bibinfo{booktitle}{\emph{Extended abstracts of the 2021
  CHI conference on human factors in computing systems}}.
  \bibinfo{pages}{1--5}.
\newblock


\bibitem[Spradling et~al\mbox{.}(2008)]%
        {spradling2008ethics}
\bibfield{author}{\bibinfo{person}{Carol Spradling}, \bibinfo{person}{Leen-Kiat
  Soh}, {and} \bibinfo{person}{Charles Ansorge}.}
  \bibinfo{year}{2008}\natexlab{}.
\newblock \showarticletitle{Ethics training and decision-making: do computer
  science programs need help?}. In \bibinfo{booktitle}{\emph{Proceedings of the
  39th SIGCSE technical symposium on Computer science education}}.
  \bibinfo{pages}{153--157}.
\newblock


\bibitem[Stavrakakis et~al\mbox{.}(2021)]%
        {stavrakakis2021teaching}
\bibfield{author}{\bibinfo{person}{Ioannis Stavrakakis},
  \bibinfo{person}{Damian Gordon}, \bibinfo{person}{Brendan Tierney},
  \bibinfo{person}{Anna Becevel}, \bibinfo{person}{Emma Murphy},
  \bibinfo{person}{Gordana Dodig-Crnkovic}, \bibinfo{person}{Radu Dobrin},
  \bibinfo{person}{Viola Schiaffonati}, \bibinfo{person}{Cristina Pereira},
  \bibinfo{person}{Svetlana Tikhonenko}, {et~al\mbox{.}}}
  \bibinfo{year}{2021}\natexlab{}.
\newblock \showarticletitle{The teaching of computer ethics on computer science
  and related degree programmes. a European survey}.
\newblock \bibinfo{journal}{\emph{International Journal of Ethics Education}}
  (\bibinfo{year}{2021}), \bibinfo{pages}{1--29}.
\newblock


\bibitem[The Moral Machine({[n.\,d.]})]%
        {moralmachine}
The Moral Machine \bibinfo{year}{[n.\,d.]}\natexlab{}.
\newblock
\newblock
\newblock
\shownote{https://www.moralmachine.net/}.


\bibitem[Trim and Gulley(2023)]%
        {10.1145/3615335.3623037}
\bibfield{author}{\bibinfo{person}{Michelle Trim} {and} \bibinfo{person}{Paige
  Gulley}.} \bibinfo{year}{2023}\natexlab{}.
\newblock \showarticletitle{Imagining, Generating, and Creating Communication
  as Feminist Pedagogical Method for Teaching Computing Ethics}. In
  \bibinfo{booktitle}{\emph{Proceedings of the 41st ACM International
  Conference on Design of Communication}} (Orlando, FL, USA)
  \emph{(\bibinfo{series}{SIGDOC '23})}. \bibinfo{publisher}{Association for
  Computing Machinery}, \bibinfo{address}{New York, NY, USA},
  \bibinfo{pages}{206–209}.
\newblock
\showISBNx{9798400703362}
\urldef\tempurl%
\url{https://doi.org/10.1145/3615335.3623037}
\showDOI{\tempurl}


\bibitem[Zuber et~al\mbox{.}(2022)]%
        {zuber2022empowered}
\bibfield{author}{\bibinfo{person}{Niina Zuber}, \bibinfo{person}{Jan Gogoll},
  \bibinfo{person}{Severin Kacianka}, \bibinfo{person}{Alexander Pretschner},
  {and} \bibinfo{person}{Julian Nida-R{\"u}melin}.}
  \bibinfo{year}{2022}\natexlab{}.
\newblock \showarticletitle{Empowered and embedded: ethics and agile
  processes}.
\newblock \bibinfo{journal}{\emph{Humanities and Social Sciences
  Communications}} \bibinfo{volume}{9}, \bibinfo{number}{1}
  (\bibinfo{year}{2022}), \bibinfo{pages}{1--13}.
\newblock


\end{thebibliography}


    \appendix

    \section{Responsible Computing Online Resources}
    \label{sec:resources}

    Below are links to some online resources for responsible computing:
\begin{enumerate}

    \item \href{https://embeddedethics.seas.harvard.edu/}{Embedded EthiCS} at Harvard University

    \item  \href{https://ethicsinsociety.stanford.edu/tech-ethics/ethics-society-and-technology-hub}{Ethics, Society, and Technology Hub} at McCoy Family Center for Ethics in Society (Stanford)

    \item The \href{https://cssh.northeastern.edu/ethics/}{Ethics Institute} at Northeastern University. Also see \href{https://vsd.ccs.neu.edu/}{Value Sensitive Design}

    \item \href{https://computing.mit.edu/cross-cutting/social-and-ethical-responsibilities-of-computing/}{Social \& Ethical Responsibilities of Computing} at MIT

    \item \href{https://computer-ethics.com/}{Computer Ethics} at UC Berkeley. Also see \href{https://data.berkeley.edu/human-contexts-and-ethics}{Human Contexts and Ethics (HCE)}

    \item Mozilla \href{https://foundation.mozilla.org/en/what-we-fund/awards/teaching-responsible-computing-playbook/}{Teaching Responsible Computing Playbook}

    \item \href{https://www.computingnarratives.com/}{Computing Ethics Narratives} at Bowdoin and Colby Colleges

    \item ACM EngageCSEdu \href{https://engage-csedu.org/ethics-and-computing/}{Embedding Ethics Repository}

    \item \href{https://sites.gatech.edu/responsiblecomputerscience/}{Responsible Computing Science} at Georgia Tech 

    \item \href{https://c4sg.cse.buffalo.edu/projects/Teaching%20Responsible%20Computing.html}{Teaching Responsible Computing} at  University at Buffalo

    \item \href{https://www.bemidjistate.edu/academics/departments/mathematics-computer-science/rcs/}{Responsible Computer Science} at Bemidji State

    \item \href{https://ethicalcs.github.io/}{Ethical CS} at Bucknell University

    \item \href{https://www.internetruleslab.com/}{Internet Rules Lab} at University of Colorado Boulder.

    \item \href{https://responsible.cs.brown.edu/}{Socially Responsible Computing} at Brown University. Also see \href{https://responsibleproblemsolving.github.io/}{Responsible Problem Solving}

    \item \href{https://csethics.allegheny.edu/}{Ethical Computer Science} at Allegheny College

    \item \href{https://ethicslab.georgetown.edu/mozilla-grant}{Ethics Lab} at Georgetown University

    \item \href{https://news.mdc.edu/role-playing-scenario-developed-at-entec/}{Role-Playing Scenario} at Miami Dade

    \item \href{https://www.scu.edu/ethics/focus-areas/technology-ethics/resources/what-we-are-doing-with-ethicalcs-at-santa-clara-university/}{Ethical CS} at Santa Clara University

    \item \href{https://www.cse.wustl.edu/~cytron/RCS/}{Responsible Computer Science} at Washington University in St. Louis
\end{enumerate}

\section{Illustrative Examples of Responsible Computing Interventions}
\label{app:examples}

\subsection{Individual Assignments or problems}

\Cref{tab:indv-prob-ex} gives examples of individual assignments/problems (which were discussed in \Cref{sec:indiv-assignment}).

\begin{table}[]
    \centering
    {\renewcommand{\arraystretch}{1.2}%
    \begin{tabular}{|P{0.1}|P{0.09}|P{0.47}|P{0.17}|P{0.17}|}
    \hline
    \textbf{Assignment} & \textbf{Relevant Course (topics)} &\textbf{Description}& \textbf{Link} &\textbf{Related}\\
    \hline
    SQL Injection attack &
Databases &
This assignment  incorporates privacy, ethics and security into introductory databases courses with an activity that lets students experiment with an SQL injection attack.  The goal of the activity is to help students understand how the attack occurs and how it can harm others.  The activity also includes discussion questions related to the ethics surrounding securing data and user privacy.
&
 Allegheny College. Created by \href{https://github.com/GatorEthics/privacy}{ Jordan Wilson, Oliver Bonham-Carter}&
Allegheny College has a collection of \href{https://csethics.allegheny.edu/}{responsible Computing activities}\\
\hline
Developers as Decision Makers &
CS 1 (conditionals) &
Students develop a scoring algorithm to determine which classmates are prioritized for housing on campus. Students use a human-centered design process to reflect on the ways in which different scoring algorithms can advantage or harm different groups of people.
& Bucknell University. Created by \href{https://ethicalcs.github.io/#decision-makers}{Evan Peck}&
Peck has a whole collection of ethical \href{https://ethicalcs.github.io/}{reflection modules for CS 1}\\ 
\hline
Elections&
Data Structures (Binary Tree and Heaps)&
These assignments focus on the U.S. elections process in the context of an introductory data structures course.&
Created by \href{https://responsibleproblemsolving.github.io/#elections}{Suresh Venkatasubramanian, Sorelle Friedler, Seny Kamara, and Kathi Fisler}&
The same group has \href{https://responsibleproblemsolving.github.io/}{other responsible computing assignments} for data structures courses\\
\hline
Thesis Advisor Allocation &
Algorithms (Stable Matching Problem)&
Assignment in this folder asks students to design a stable allocation of students to thesis advisors and asks them to reflect on the implications of the their design choices.&
Haverford College. Created by \href{https://github.com/responsibleproblemsolving/algorithmdesign/tree/master/Chapter1_Introduction_Some_Representative_Problems/thesis_advisor_allocation}{Sorelle Friedler}&
This \href{https://github.com/responsibleproblemsolving/algorithmdesign/tree/master}{github page has collection of other responsible algorithms problems}\\
\hline
College Admissions Algorithms&
CS 1 (lists and functions)&
Create an automated system to process college applications&
University of Colorado Boulder. Created by \href{https://www.internetruleslab.com/ethicsbased-computer-science-assignments}{Natalie Garrett and Casey Fiesler}&
There are other \href{https://www.internetruleslab.com/ethicsbased-computer-science-assignments}{CS 1 assignments} from the same group\\
\hline
Algorithmic Fairness&
Algorithms&
Ausitn's cake cutting algorithm as an example of an inherently fair algorithm. Cake cutting algorithms consider the fair division of a resource that can be divided without losing value. Austin's algorithm guarantees a fair outcome even if the parties try to cheat&
University of Washington in St. Louis. Created by \href{https://www.cse.wustl.edu/~cytron/RCS/}{Ron K. Cytron, Tomas Larsen, Russell Scharf}&
The same group has \href{https://www.cse.wustl.edu/~cytron/RCS/}{other activities}\\
\hline
    \end{tabular}
    }
    \caption{Examples of Individual Assignments or problems.}
    \label{tab:indv-prob-ex}
\end{table}

\subsection{Self-contained Lesson/Module}

\Cref{tab:self-cont-module-ex} gives examples of self-contained lessons/modules (which were discussed in \Cref{sec:self-contained-module}).

\begin{table}[]
    \centering
    {\renewcommand{\arraystretch}{1.2}%
    \begin{tabular}{|P{0.1}|P{0.09}|P{0.47}|P{0.17}|P{0.17}|}
    \hline
    \textbf{Module} & \textbf{Courses} &\textbf{Description}& \textbf{Link} &\textbf{Related}\\
    \hline
Speculative Ethics Classroom Exercises&
Multiple&
This page contains information and resources about the Black Mirror Writers Room and teaching exercises for ethical speculation in computing.&
University of Colorado Boulder. Created by \href{https://www.internetruleslab.com/black-mirror-writers-room}{Casey Fiesler}&
\\
\hline
Ethical Implications of the Adoption of Facial
Recognition Technology role play&
Multiple&
Introduces  social responsibility via an activity that simulates a city hall discussion about adopting facial recognition technology,  The overall goal is to encourage students to evaluate the values of decision makers vs. those affected in the community as it relates to the adoption of technology.  The materials for the role play include a teaching plan, case, and discussion prompts&
Miami Dade College. \href{https://www.mdc.edu/entec/downloads/rpg-facial-recognition.pdf}{Ethical Implications of the Adoption of Facial Recognition Technology. School of Engineering and Technology}&
\\
\hline
An Introduction to Data Ethics&
Data Science courses&
This introductory ethics module for data science courses includes a reading, homework assignments, and case studies, all designed to spark a conversation about ethical issues that students will face in their role as data practitioners
&
Markkula Center for Applied Ethics, Santa Clara University. Created by \href{https://www.scu.edu/ethics/focus-areas/technology-ethics/resources/an-introduction-to-data-ethics/}{Shannon Vallor.}&
Markkula Center has other \href{https://www.scu.edu/ethics/focus-areas/internet-ethics/teaching-modules/ and more general resources: https://www.scu.edu/ethics/focus-areas/technology-ethics/resources/embedding-ethics-into-computing-curricula-resources-and-suggestions/}{resources on tech and ethics}
\\
\hline
A City Decides On Self-Driving Buses&
Multiple&
Student do role play as different stakeholder in a decision at the city level on whether the city will allow self driving cars or not&
Georgia Institute of Technology. Created by \href{https://sites.gatech.edu/responsiblecomputerscience/overview/}{Ellen Zegura, Jason Borenstein, Benjamin Shapiro, Amanda Meng, and Emma Logevall}&
The same group \href{https://sites.gatech.edu/responsiblecomputerscience/}{other role playing activities}
\\
\hline
Making Computing Anti-Racist&
First year seminar&
First-year University at Buffalo students accepted the challenge to spend two weeks of their semester imagining what it would take to build a world in which computing could become anti-racist. Starting with the specific case of the use of predictive policing algorithms, they proposed computational and non-computational solutions to the problems exacerbated by technology in society&
University at Buffalo. Created by \href{https://www-student.cse.buffalo.edu/cseimpossibleproject/}{Kenneth Joseph and Dalia Muller}&
\\
\hline
Crypto and Cypherpunk Ethics&
Security/ Cryptography&
This module draws upon cypherpunk and cypherpunk-related ethical analyses of cryptography to explore the ongoing debates involving personal privacy, national security, system/device security and the meaning of an open society. Through reading, discussion and small group work, students will develop conceptual and practical knowledge about the ethics of cryptography&
\href{https://www.bemidjistate.edu/academics/departments/mathematics-computer-science/rcs/teaching-modules/crypto-and-cypherpunk-ethics/}{
Bemidji State University}&
The group has other material related to \href{https://www.bemidjistate.edu/academics/departments/mathematics-computer-science/rcs/teaching-modules/}{teaching responsible CS}
\\
\hline
Value Sensitive Design&
Multiple&
Value Sensitive Design, or VSD, is a framework for integrating ethics and values into the design of technological systems. VSD is the unifying foundation for a interdisciplinary, College-wide effort to integrate ethics into undergraduate education in the Khoury College. &
\href{https://vsd.ccs.neu.edu/}{Northeastern University}
&\\
\hline
    \end{tabular}
    }
    \caption{Examples of Self-contained Lesson/Module.}
    \label{tab:self-cont-module-ex}
\end{table}

\subsection{Integrated Lesson/Module}

\Cref{tab:int-module-ex} gives examples of integrated lessons/modules (which were discussed in \Cref{sec:integrated-module}).

\begin{table}[]
    \centering
    {\renewcommand{\arraystretch}{1.2}%
    \begin{tabular}{|P{0.1}|P{0.09}|P{0.47}|P{0.17}|P{0.17}|}
    \hline
    \textbf{Name} & \textbf{Courses} &\textbf{Description}& \textbf{Link} &\textbf{Related}\\
    \hline
Inclusive HCI&
Human Computer Interaction (HCI)&
The aim of this exercise is to dispel the myth of the universal user and increase students’ awareness of accessibility challenges (and opportunities!) in human-computer interactions. Consisting of an individual homework assignment and an in-class group activity, it is meant to enrich and activate students’ awareness of exclusion in design and help them think through important philosophical issues in inclusive design.&
Georgetown University. Created by \href{https://ethicslab.georgetown.edu/blog/responsible-cs-exercise-inclusive-hci-now-available}{Ethics Lab}
&Ethics Labs has other \href{https://ethicslab.georgetown.edu/mozilla-grant}{related resources}\\
\hline
Embedded Ethics&
Integrated Lesson/Module&
Embeds “philosophers directly into computer science courses” through a repository of modules that include cases, teaching plans for discussing relevant ethical issues in various CS courses.  The Embedded Ethics project provides SEP curricula for numerous CS courses and can be adopted to provide pedagogy on ethics through an undergraduate program&
Harvard University. Created by \href{https://embeddedethics.seas.harvard.edu/}{Embedded EthiCS}
&Other schools now have similar programs now, E.g. at \href{https://embeddedethics.stanford.edu/}{Stanford University} 
\\
\hline
Building ethical guardrails for Technological products/artifacts from a stakeholder perspective&
Multiple &
A project that builds on decolonial and feminist pedagogical practices. Works as a project or assignment for graduate and undergraduate students. Works best if students have been exposed to literature or examples of algorithmic bias, exclusionary system or interface design, and/or social impacts from computing technology.&
U. Mass Amherst. Created by Michelle D. Trim and Paige Gulley~\cite{10.1145/3615335.3623037}&
\\
\hline

\hline
    \end{tabular}
    }
    \caption{Examples of Integrated Lesson/Module.}
    \label{tab:int-module-ex}
\end{table}

\subsection{Responsible Computing Theme}

\Cref{tab:rc-theme-ex} gives examples of responsible computing themes (which were discussed in \Cref{sec:RC-theme}).

\begin{table}[]
    \centering
    {\renewcommand{\arraystretch}{1.2}%
    \begin{tabular}{|P{0.1}|P{0.09}|P{0.47}|P{0.22}|P{0.12}|}
    \hline
    \textbf{Name} & \textbf{Courses} &\textbf{Description}& \textbf{Link} &\textbf{Related}\\
    \hline
Human Context and Ethics Toolkit&
Multiple&
A set of concepts and methods from Science, Technology, and Society (STS) and History selected to build understanding of the datafied world, helping students to identify where human power structures and value choices get built into technical work, and empowering them to discover how, when, and where they can responsibly and effectively intervene.&
University at California, Berkeley. Created by \href{https://data.berkeley.edu/hce-toolkit}{Margarita Boenig-Liptsin, Cathryn Carson} 
&\href{https://data.berkeley.edu/human-contexts-and-ethics}{Human Contexts and Ethics Program at Berkeley}\\ 
\hline
Access to High Speed Internet&
Algorithms&
There are multiple activities/assignments in CSE 331 that all have the common thread of accessing high speed Internet.&
University at Buffalo. Created by \href{https://c4sg.cse.buffalo.edu/projects/Teaching%20Responsible%20Computing.html#algorithms-course-cse-331}{Atri Rudra, CSE 331 (Algorithms and Complexity)}.&
\\
\hline
    \end{tabular}
    }
    \caption{Examples of Responsible Computing Themes.}
    \label{tab:rc-theme-ex}
\end{table}

\subsection{Dedicated Course}

\Cref{tab:dedicated-ex} gives examples of responsible computing themes (which were discussed in \Cref{sec:dedicated}).

\begin{table}[]
    \centering
    {\renewcommand{\arraystretch}{1.2}%
    \begin{tabular}{|P{0.1}|P{0.1}|P{0.6}|P{0.2}|}
    \hline
    \textbf{Course} & \textbf{Required vs. Elective} &\textbf{Description}& \textbf{Link}\\
    \hline
Algorithms for the People&
Elective&
The course explores the ways in which technology affects marginalized communities addressing issues such as geo-fencing, policing, human rights, and financial technologies.  The course is offered among electives for students concentrating in CS.  Algorithms for the People include assigned readings on course themes and a project that evaluates existing solutions or develop and analyze new algorithms that affect marginalized populations.&
Brown University. Created by Seny Kamara, \href{http://algosforthepeople.org/}{2950v-19. “Algorithms for the People.” Algorithms for the People, 26 June 2020}\\
\hline
Philosophy and Society and Ethics and Information Technology&
Required Elective&
Students are required to select among several SEP courses, including  Ethics; Philosophy and Society, and Ethics and Information Technology which are taught in the department of Philosophy.  The Ethics and Philosophy and Society courses cover traditional theories and analysis of applied ethics, while the Ethics and Information Technology course addresses “contemporary ethical issues concerning the use, misuse and development of information technologies”&
University of Colorado at Boulder. \href{https://www.colorado.edu/cs/academics/undergraduate-programs/bachelor-science/bachelor-science-degree-requirements#Logic}{Bachelor of Science Degree Requirements at UC Boulder}.
\\
Social and Ethical Issues in Information Technology&
Required&
incorporates ethical reasoning and the ACM Code of Ethics to examine a wide range of issues including privacy, security and encryption, intellectual property, censorship and computer crime. &
University of Maryland Baltimore County. \href{https://www.csee.umbc.edu/cmsc-304-syllabus/}{CMSC 304 Syllabus}
\\
\hline
Race, Gender, Class, \& Computing &
Elective&
This course explores the diversity, equity, and inclusion (DEI) challenges in computing through an introduction to and discussion of identity as a social construct, its impact on computing departments and organizations, and the resulting impact of technology on various identities.&
Duke. Created by \href{https://courses.cs.duke.edu/fall21/compsci240/}{Nicki Washington}
\\
\hline
Rage Against the Machine + Machine Learning and Society&
Elective&
Two concurrently run courses: Rage Against the Machine (History) and Machine Learning and Society (CSE) . Rage Against the Machine is a class that explores the history of white supremacy in and beyond the United States. Machine Learning and Society talks about the interaction of the ML pipeline with society. Both courses  had a common project (with team members from both courses) on “Ending White Supremacy”&
University at Buffalo. Dale Muller: \href{https://daliamul.wixsite.com/rage}{Rage Against the Machine}, Kenneth Joseph + Atri Rudra \href{http://www-student.cse.buffalo.edu/~atri/ml-and-soc/spr23/}{ML and Society}
\\
\hline
    \end{tabular}
    }
    \caption{Examples of Dedicated courses.}
    \label{tab:dedicated-ex}
\end{table}

\end{document}